\documentclass[usenatbib]{mn2e}
\usepackage[dvips]{graphicx}
\usepackage[dvips]{rotating}
\usepackage{epsfig}
\usepackage{lscape}
\usepackage{amssymb,amsmath}

\newcommand{\kms}{{\rm \,km\,s^{-1}}}
\newcommand{\Mpc}{\,{\rm Mpc}}
\newcommand{\kpc}{\,{\rm kpc}}
\newcommand{\Reff}{r_{\rm e}}
\newcommand{\fR}{f_{\rm re}}
\newcommand{\fcool}{f_\star}
\newcommand{\CritDens}{\Sigma_{\rm cr}}
\newcommand{\Dl}{D_{\rm d}}
\newcommand{\Ds}{D_{\rm s}}
\newcommand{\Dls}{D_{\rm ds}}
\newcommand{\tableline}{{}}
\newcommand{\Nsub}{N_{\rm sub}}
\newcommand{\fsub}{f_{\rm sub}}
\newcommand{\msub}{m_{\rm sub}}
\newcommand{\hsml}{{\cal{H}}}
\newcommand{\Rcusp}{R_{\rm cusp}}

\newcommand{\araa}{Annual Review of Astronomy and Astrophysics}
\newcommand{\mnras}{MNRAS}
\newcommand{\apjl}{ApJ Letters}
\newcommand{\apjs}{ApJ Suppl.}
\newcommand{\apj}{ApJ}
\newcommand{\aj}{AJ}
\newcommand{\nat}{Nature}
\newcommand{\aap}{A\&A}

\title{Effects of Dark Matter Substructures on Gravitational Lensing:
  Results from the Aquarius Simulations} \author[Xu et al.]
{D. D. Xu$^{1}$\thanks{E-mail: Dandan.Xu@postgrad.manchester.ac.uk}, Shude
  Mao$^{1}$, Jie Wang$^{2,3}$, V. Springel$^{2}$, Liang Gao$^{3,4}$,
  S.D.M. White$^{2}$, \and Carlos S. Frenk$^{3}$, Adrian Jenkins$^{3}$,
  Guoliang Li$^{5}$, Julio  F. Navarro$^{6}$\\
  $^{1}$ Jodrell Bank Centre for Astrophysics, the University of Manchester,
  Alan Turing Building, Manchester M13 9PL, United Kingdom \\
  $^{2}$ Max-Planck Institut F\"ur Astrophysik, Karl-Schwarzshild-Stra{\ss}e 1,
  85740 Garching, Germany \\
  $^{3}$ Institute of Computational Cosmology, Dept. of Physics,
  University of Durham, South Road, Durham DH1 3LE, United Kingdom \\
  $^{4}$ National Astronomical Observatories, Chinese Academy of Sciences,
  Beijing, 100012, China \\
  $^{5}$ Argelander-Institut f\"ur Astronomie, University of Bonn,
  Auf dem H\"ugel 71, D-53121 Bonn, Germany \\
  $^{6}$ Department of Physics and Astronomy, University of Victoria,
  Victoria, BC, V8P, 5C2, Canada }

\begin{document}
\date{Accepted ...... Received ...... ; in original form......   }

\pagerange{\pageref{firstpage}--\pageref{lastpage}} \pubyear{2002}
\maketitle
\label{firstpage}
\begin{abstract}
  We use the high-resolution Aquarius simulations of the formation 
  of Milky Way-sized haloes in the $\Lambda$CDM cosmology to study the 
  effects of dark matter substructures on gravitational lensing.
  Each halo is resolved with $\sim 10^8$ particles
  (at a mass resolution $m_{\rm p} \sim 10^3$ to $10^4 h^{-1}M_{\odot}$)
  within its virial radius. Subhaloes with masses
  $ \msub \ga 10^5 h^{-1}M_{\odot}$ are well resolved,
  an improvement of at least two orders of magnitude over
  previous lensing studies. We incorporate a baryonic component modelled as a
  Hernquist profile and account for the response of the dark matter via
  adiabatic contraction.
We focus on the ``anomalous'' flux ratio problem, in particular on
the violation of the cusp-caustic relation due to substructures.
We find that subhaloes with masses less than $\sim 10^8
h^{-1}M_{\odot}$ play an important role in causing flux anomalies;
such low mass subhaloes have been unresolved in previous studies.
There is large scatter in the predicted flux ratios between different
haloes and between different projections of the same halo. 
In some cases, the frequency of predicted anomalous flux ratios is 
comparable to that observed for the radio lenses, 
although in most cases it is not.  
The probability for the simulations to reproduce the observed violations 
of the cusp lenses is $\approx 10^{-3}$. We therefore conclude that 
the amount of substructure in the central regions of the Aquarius haloes 
is insufficient to explain the observed frequency of violations 
of the cusp-caustic relation. 
These conclusions are based purely on our dark matter simulations 
which ignore the effect of baryons on subhalo survivability.

\end{abstract}


\section{INTRODUCTION}

Currently there are $\sim 200$ known galaxy-scale
lenses, divided roughly equally in number into lensed active galactic
nuclei\footnote{http://www.cfa.harvard.edu/castles/} and lensed background
galaxies (\citealt{Bolton2008}). These galaxy-scale lenses allow diverse
applications (see the review ``Strong Gravitational Lensing'' 
by Kochanek in \citealt{Kochanek06}) such as a determination of the 
Hubble constant, a characterisation of galaxy evolution,
and measurements of the mass distribution in galaxies. The last application
will likely be the most important one in the next decade, since there are few
other probes at intermediate redshifts ($z \sim 0.5-1$).

It was noticed quite early on that the flux ratios of the
multi-lensed images are more difficult to reproduce with simple
parametric mass models than the image positions
(\citealt{Kochanek91}). This has been termed the ``anomalous flux
ratio'' problem. Image positions and magnifications (flux ratios)
are determined by the first-order and second-order derivatives of
the lensing potential, respectively. Therefore flux ratios, as a
high-order derivative, are expected to be more sensitive to small
changes in the lensing potential than image positions.

In this regard, gravitational lenses with two or three close images
deserve special attention because, in these cases, the sources must
be close to either a fold or a cusp of the caustic.  It is well
known for any smooth lensing potential that the close images follow
{\it asymptotic} flux ratio relations: for a close pair, their flux
ratio approaches unity when their separation goes to zero, while for
a close triple, the ratio of the flux of the middle image to the sum
of the fluxes of the two outer images asymptotically goes to unity
(\citealt{Mao92, SW1992aa, KGP2003apj}; \citealt{CKN2008}). However,
the observed lensing systems often violate these asymptotic
relations. This was taken to be evidence for substructure in lensing
galaxies (\citealt{MS1998mn, MM2001, MZ2002, DK2002, Chiba2002,
KD2004}) on the physical scale of the separation between close
images (typically of the order of $\sim$ 1 kpc).
Spectroscopic observations go beyond simple
broad-band flux ratios and provide a promising way to probe
substructure in lenses (\citealt{Metcalf2004, Chiba2005, Sugai2007}).
Other suggestive evidence for substructures comes from astrometry (see
\citealt{Chen2007} for a general discussion), such as bent jets
(\citealt{Metcalf2002}), and detailed image structures for B2016+112
(\citealt{Kochanek06, Koopmans2002, More2009}) and B0128+437
(\citealt{Biggs2004, Zhang2008}). Substructures may also have
detectable effects on the time delays in gravitational lenses
(\citealt{KM2008}).

\citet{EW2003} argued that some of these lensing ``anomalies'' 
may be accommodated by changes in the potentials of the main lensing 
galaxies in parametric models. However, significant changes are needed 
in order to explain the anomalies (\citealt{KD2004,CongdonKeeton2005}).
The angular structures of the lenses whenever the measurements
are available suggest ellipsoidal central potentials, where the high
amplitude, higher order multipoles that are required to explain the flux
ratio anomalies are not seen (\citealt{KD2004,Yoo2005,Yoo2006}). 
There are strong hints that substructures may indeed be real
in lensing galaxies. First, observationally, saddle (negative-parity)
images are often fainter than the predictions of simple smooth models.
This is expected from lensing by substructure (such as stars, or subhaloes;
\citealt{SW2002apj}; \citealt{KD2004}), but impossible to explain by
propagation effects, such as galactic scintillation and scatter
broadening, as earlier postulated (\citealt{Koopmans2003JVASCLASS}).
This arguably constitutes the most convincing evidence for
substructure lensing. Second, in many gravitational lenses, the
substructure is directly seen as luminous satellites. For example,
nearly half of the CLASS lenses (\citealt{Browne2003, Myers2003,
Jackson2009}) show luminous satellite galaxies within a few kpc of
the primary lensing galaxies\footnote{This fraction
  is a factor of $\sim 2$ higher than that claimed in \citet{BMK2008}
  as revealed by a more careful analysis of HST images of the CLASS lenses
  (\citealt{Jackson2009}).}.
Inclusion of satellites in the modelling dramatically improves
the fit to the image positions. In the case of B2045+265,
the inclusion of a companion galaxy helps to explain the flux ratio
anomaly (\citealt{McKean2007}). 
The additional dark subhaloes within the main lensing galaxies, 
as well as the intergalactic perturbers along the line-of-sight 
(\citealt{Chen2003, Wambsganss2005, Metcalf2005a}a,b; \citealt{Miranda2007})
may also help to explain the observed lensing anomalies.

Much of the interest in (milli-)lensing flux anomalies arises because
they may be caused by the elusive substructure generically predicted by 
the hierarchical structure formation in the cold dark matter (CDM) cosmology
(e.g. \citealt{Kauffmann1993, Klypin1999apj, Moore1999apj, Ghigna2000,
  Gao2004a}a,b; \citealt{Diemand2007apj}). In this model, large structures
form via merging and accretion of smaller structures. The cores of
these small structures often survive tidal destruction and manifest
themselves as subhaloes (substructure). Recent high-resolution
simulations predict many thousands of subhaloes (down to $\msub \sim
10^6 M_\odot$, or to circular velocity of $V_{\rm c} \sim 4 \kms$;
e.g.  \citealt{MDK2008, volker08Aq}), at least two orders
of magnitude more than the number of observed satellite galaxies in
the Milky Way, even after accounting for the newly discovered faint
satellite galaxies from the Sloan Digital Sky Survey
(\citealt{Belokurov2007}). A possible solution is that star formation
may be strongly suppressed in the vast majority of the low-mass
subhaloes (e.g. \citealt{Efstathiou1992, Kauffmann1993,TW1996,
Bullock2000,Gnedin2000,Benson2002}), and thus they remain dark and
difficult to detect through light-based methods. If this is the case,
then gravitational lensing can potentially probe this population since
it depends only on the mass but not on whether the lenses are luminous
or dark.

Numerical simulations indicate that subhaloes typically account for 5-10 per
cent of the total mass in a galaxy-type halo (e.g.  \citealt{Klypin1999apj,
  Moore1999apj, Ghigna2000}). The study by \citet{DK2002} requires $\fsub =$
0.6\% to 7\% (with a median of 2\%) of the mass to be in
substructures (90\% confidence limit) in order to explain the observed
flux anomaly problem. At first sight, the fraction of substructure from
simulations seems to be more than sufficient to explain the flux
anomaly. Upon closer examination, however, a problem emerges:
lensing probes the central few kpc around the line-of-sight through
the galaxy, while most substructures are in the outer regions of its
dark matter halo, since those that come close to the centre are
tidally destroyed. Thus it remains unclear whether the predicted
substructure in the inner regions is sufficient or not to explain
the observed flux anomalies (e.g. \citealt{BS2004aa,
MaoJing04apj, Maccio2006, AB06mn}). In contrast, on cluster scales,
the amount of predicted substructure seems to be consistent with
weak and strong lensing data (\citealt{Natarajan2007}).

Previous lensing studies simulated galaxy-sized haloes
with $\sim 10^6$ particles so that subhaloes were resolved down
to $\sim 10^7$ to $10^8 h^{-1}M_{\odot}$.
State-of-the-art simulations can now resolve haloes with two
or even three orders of magnitude more particles, thus reaching substantially
lower mass subhaloes. In this work, we revisit the issue of substructure
lensing using the Aquarius simulations of six galaxy-sized haloes.  These
collisionless $N$-body simulations were performed by the Virgo Consortium in a
concordance $\Lambda$CDM universe. The subhaloes in each halo are resolved
down to masses of $\msub \sim 10^5 h^{-1}M_{\odot}$ (\citealt{volker08Aq}), at
least two orders of magnitude better than that in previous
substructure lensing studies.

Our paper is organised as follows. In Section 2, we describe the realisation
and the properties of the simulated lensing galaxies. Our methods and
techniques for the lensing simulations together with our test results are
presented in Section 3. In Section 4, we apply our lensing simulation to the
six simulated galaxy haloes from the Aquarius simulation to derive their
lensing properties, including the cusp relations, and we compare the numerical
results with observations. A summary of the paper and a discussion are given
in Section 5. The cosmology we adopt for the lensing simulation is the same as
that used for the Aquarius simulations (\citealt{Volker05Nature}), with a
matter density $\Omega_{\rm m}$ = 0.25, cosmological constant
$\Omega_{\Lambda}$ = 0.75, Hubble constant $h=H_0/(100\kms\Mpc^{-1})=0.73$
and linear fluctuation amplitude $\sigma_8=0.9$.

\section {From Dark Matter haloes to Early-Type Lensing Galaxies }
\label{sec:darkLight}

In this section, we summarise the properties of dark matter haloes from the
Aquarius simulations relevant to our study, in particular the subhalo
properties. Readers are referred to \citet{volker08Aq} for more details. We
will show that dark matter alone is, as expected, insufficient to cause
multiple image splittings, and therefore we must incorporate a stellar
component; we detail such a procedure in \S\ref{sec:insertBaryon}.

\subsection{The Aquarius simulations}

The Aquarius project (\citealt{volker08Aq}) is a suite of simulations of six
galaxy-sized dark matter haloes with five levels of numerical resolution. The
haloes were selected from a 100$h^{-1}$ Mpc simulation box within the
concordance cosmology (for parameters see above). The simulations were run
with GADGET-3, an improved version of the GADGET-2 code
(\citealt{VolkerGADGET1, VolkerGADGET2}). The highest resolution level (level
1) was achieved for only one halo (``{\it Aq-A-1}'') with $\sim 1.5$ billion
halo particles. Level-2 simulations were performed for a sample of six
dark matter haloes, with about 200 million particles per halo. The softening
length is $\sim$0.05$h^{-1}$ kpc, and the mass resolution ranges
from $10^3$ to $10^4 h^{-1}M_\odot$.
All haloes are Milky-Way type systems in terms of their mass
and rotation curve. We will use the six level-2 haloes ({\it Aq-A-2}, {\it
  Aq-B-2}, {\it Aq-C-2}, {\it Aq-D-2}, {\it Aq-E-2} and {\it Aq-F-2}) at
redshift zero for our analysis of substructure lensing. As we will show later
on, the scatter in lensing properties among different haloes (and for
different projections) is large, and so it is important to examine more than
one halo for statistical purposes.

The basic properties of the six haloes at $z=0$ are listed in Table
\ref{tab:Aquarius_simulations}.  In particular, all the density profiles are
reasonably fit by Navarro, Frenk and White (NFW) profiles (\citealt{NFW96,
  NFW97})\footnote{An even better fit is found using the
  \citet{Einasto66} profile (\citealt{Navarro08mp}), but here we adopt the
  simpler NFW profile which we use later to take into account the adiabatic
  contraction of dark matter haloes.}:
\begin{equation}
\begin{array} {c}
\displaystyle \rho(r) = \frac{M_{200}}{4\pi r (r+r_{200}/c)^2 f(c)},\\
\displaystyle M(<r) = \frac{M_{200}f(r\,c/r_{200})}{f(c)},\\
\displaystyle f(c) = \ln(1+c)-c/(1+c),
\end{array}
\label{eq:NFW}
\end{equation}
where $r_{200}$ is the radius within which the mean dark halo mass density is
200 times the critical density, $M_{200}$ is the mass enclosed within
$r_{200}$, and $c \equiv r_{200}/r_{\rm s}$ is the concentration parameter
with $r_{\rm s}$ being the scale radius.

\begin{table*}
\centering
\caption{Dark matter halo properties in the Aquarius simulations:}
\label{tab:Aquarius_simulations}
\begin{minipage} {\textwidth}
\begin{tabular}[b]{c|c|c|c|c|c|c|c}\hline
Halo Name & $r_{200}$ & $M_{tot}$ & $c$ & Mass Resolution & $N_{200}$ & $\Nsub$ & $\fsub$ \\
~ & ($h^{-1}$ kpc) & ($10^{10} h^{-1}M_{\odot}$) & ~& ($h^{-1}M_{\odot}$) & ~ & ~ & (per cent) \\\hline
Aq-A-2 & 179.5 & 132.8 & 16.2 & $1.0 \times 10^4$ & $1.3 \times 10^8$ & $2.1 \times 10^4$ & 7.14 \\
Aq-B-2 & 137.1 & 59.5 & 9.7 & $4.7 \times 10^3$ & $1.3 \times 10^8$ & $2.5 \times 10^4$ & 6.98 \\
Aq-C-2 & 177.3 & 127.7 &  15.2 & $1.0 \times 10^4$ & $1.2 \times 10^8$ & $1.7 \times 10^4$ & 4.12 \\
Aq-D-2 & 177.3 & 128.5 & 9.4 & $1.0 \times 10^4$ & $1.3 \times 10^8$ & $2.2 \times 10^4$ & 6.56 \\
Aq-E-2 & 155.0 & 85.7 & 8.3 & $7.0 \times 10^3$ & $1.2 \times 10^8$ & $2.3 \times 10^4$ & 7.28 \\
Aq-F-2 & 153.0 & 80.5 & 9.8 & $4.9 \times 10^3$ & $1.6 \times 10^8$ & $2.6 \times 10^4$ & 11.20 \\
Aq-A-2 ($Z$ = 0.6) & 134.4 & 92.2 & 10.4 & $1.0 \times 10^4$ & $9.3 \times 10^7$ & $1.7 \times 10^4$ & 6.50 \\
\hline \\
\end{tabular}
\\
Note: Col (1): halo name, Cols (2)-(4): $r_{200}$, $c$ and $M_{200}$ are
defined in eq. (\ref{eq:NFW}) for the main halo. $M_{\rm tot}=M_{200}+M_{\rm
  sub}$, where $M_{\rm tot}$ and $M_{\rm sub}$ are the total masses of all
dark matter and of all the subhaloes within $r_{200}$.  Col (5): Mass
resolution ($h^{-1}M_{\odot}$).  Col (6): $N_{200}$ is the total number of
particles within $r_{200}$.  Col (7): $\Nsub$ is the number of subhaloes
within $r_{200}$.  Col (8): $\fsub$ is the mass fraction of subhaloes within
$r_{200}$, defined by $M_{\rm sub}/M_{\rm tot}$. \\
\end{minipage}
\end{table*}

\begin{figure}
\centering
\includegraphics[width=8cm]{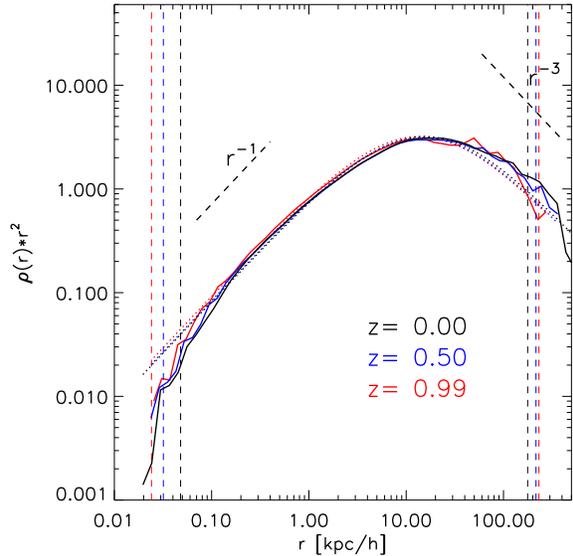}
\caption{Density profiles (solid curves), multiplied by $r^2$,
  for the halo {\it Aq-D-2} at
  redshifts $z=0$, 0.5 and 0.99.  All haloes are reasonably well fitted by the
  NFW profile (dotted curves, see Eq.~\ref{eq:NFW}), which follows
  $\rho(r)\propto r^{-1}$ on small scales and $\rho(r)\propto r^{-3}$ on large
  scales. The vertical dashed lines indicate the softening length and
  $r_{200}$.}
\label{fig:Mainhalo1}
\end{figure}

We artificially put all these haloes (snapshot $z=0.0$) at redshift $z=0.6$
(corresponding roughly to the most likely lens redshift, e.g.  \citealt{TOG}),
keeping their physical sizes unchanged. However, we also take a
snapshot of the halo {\it Aq-A-2} at redshift $z=0.6$ as a lens, and compare
its lensing properties with those artificially shifted to $z=0.6$. As will be
shown in \S4, the scatter among the six haloes is much larger than the
differences between haloes at redshifts $z=0$ and $z=0.6$, and so adopting the
$z=0$ haloes will not significantly change the properties of substructure
lensing.  This is also seen in the evolution of density profiles of these
haloes.  Fig.~\ref{fig:Mainhalo1} shows the density profiles for the halo {\it
Aq-D-2} at redshifts 0, 0.50 and 0.99.
The changes in the profiles since
redshift~1 are relatively small since the Aquarius
haloes form earlier than that.

As we are primarily interested in the substructure lensing, an important step
is the identification of the subhaloes. We use the {\tt SUBFIND} routine
(\citealt{VolkerGADGET2}) to identify subhaloes exceeding 20 particles, which
corresponds to a minimum subhalo mass of $\sim 10^5 h^{-1}M_{\odot}$. The
number of subhaloes in each halo ranges from about $1.7\times 10^4$ to
$2.6\times 10^4$ within $r_{200}$, with 4.1-11.2 per cent of the total halo
mass locked up in bound subhaloes (see Col (8):
$\fsub$ in Table \ref{tab:Aquarius_simulations}).

The subhalo mass function follows a power-law:
${\rm d}N(\msub)/{\rm d}\msub \propto \msub^{-1.9}$ (\citealt{volker08Aq}).
The average mass of subhaloes (within $r_{200}$)
is $\sim 10^6$ to $10^7 h^{-1}M_{\odot}$ and their average
half-mass radius is $\le 0.2h^{-1}$ kpc, with large scatter.  The most massive
subhalo has a mass of $\sim 10^9$ to $10^{10} h^{-1}M_{\odot}$
and a half-mass radius $\sim 5-10 h^{-1}$ kpc.

\begin{figure*}
\centering
\includegraphics[scale = 0.45]{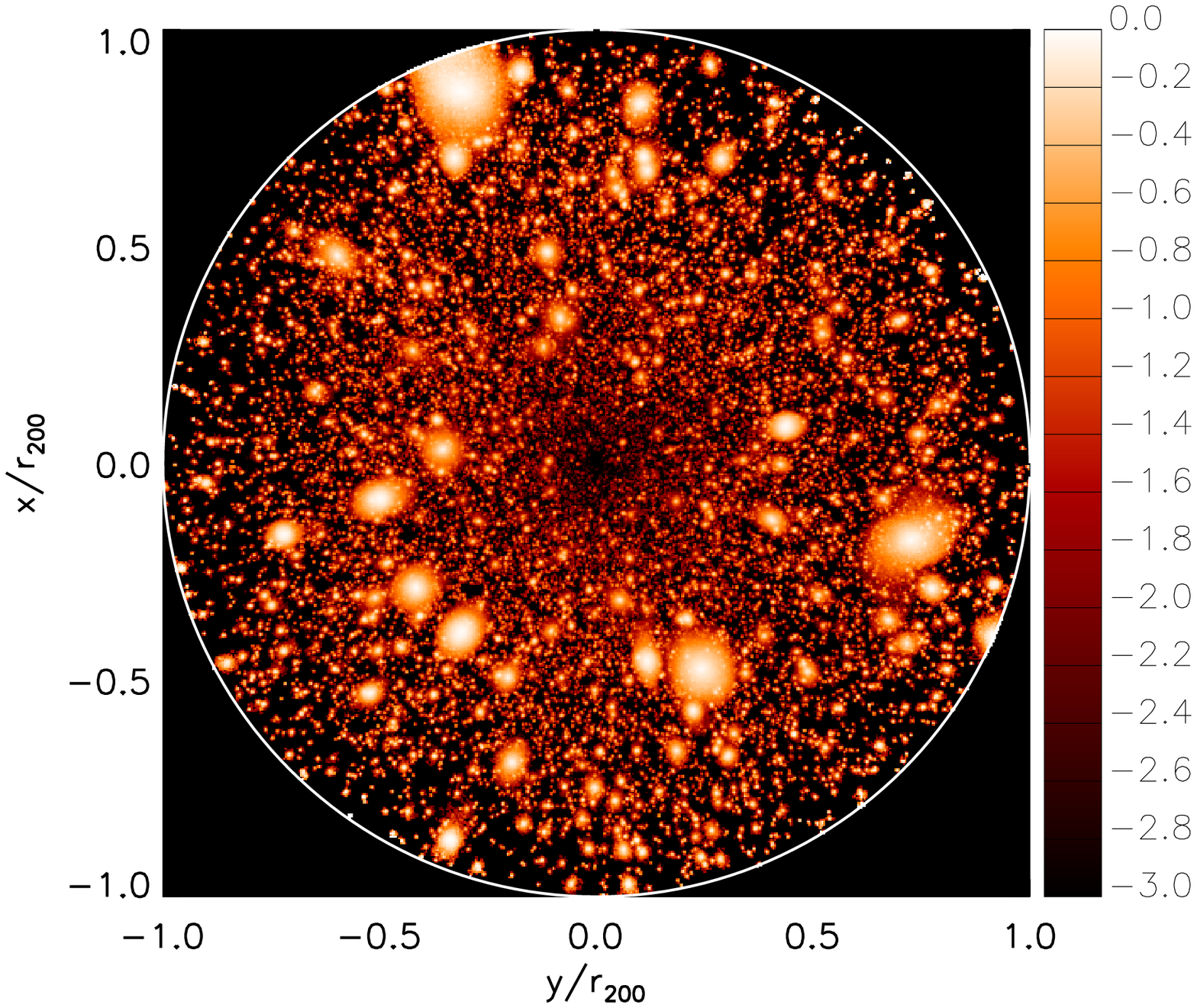}
\includegraphics[scale = 0.52]{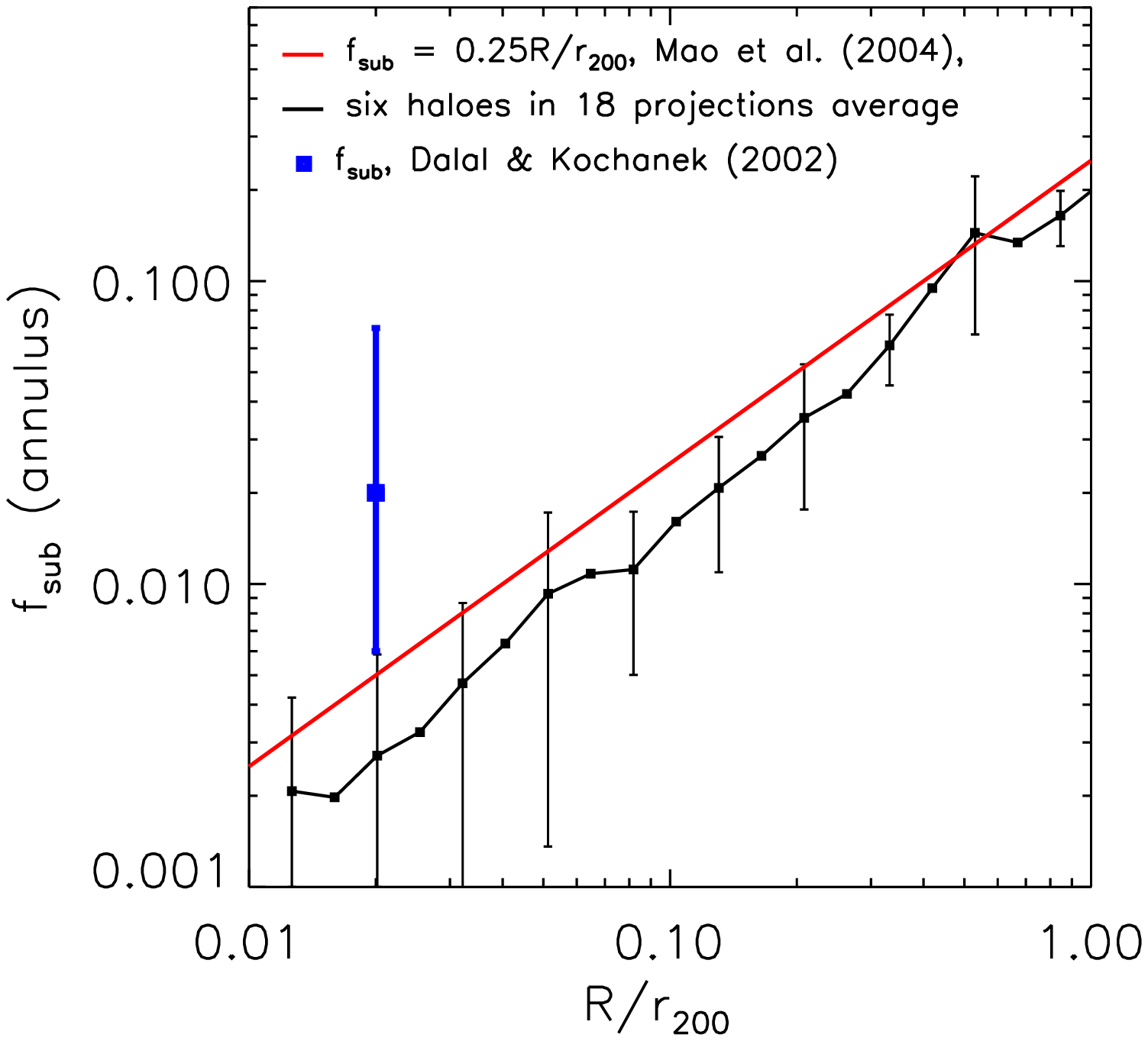}
\caption{The left panel shows a contour map of the subhalo
surface mass density fraction, which is the ratio of the surface
mass in subhaloes to that in the total halo, for {\it Aq-D-2}
projected along the $Y$-axis. The right panel shows the mean
  distribution of subhalo surface mass fraction as a function of $R/r_{200}$,
  averaged over the three independent projections of each of the six
  Aquarius haloes at redshift $z=0$. The error bars indicate the 68\%
  scatter among different projections and haloes. The red lines show
  the fit from \citet{MaoJing04apj}. The blue point indicates
  the median and 90\% confidence level of the required fraction found 
  by \citet{DK2002} (assuming the Einstein radius to be 0.02 $r_{200}$).}
\label{fig:Subhalo_properties}
\end{figure*}

As an example, we again consider halo {\it Aq-D-2} and show in the left panel
of Fig.~\ref{fig:Subhalo_properties} the $Y$-projection of the surface mass
fraction in subhaloes within $r_{200}$.
The right panel in the same figure shows the surface mass fraction of
subhaloes averaged within azimuthal annuli
as a function of the normalised radius $R/r_{200}$. It is clear that the
scatter in the projected mass fraction of subhaloes among different haloes is
large. Within 0.1 $r_{200}$, the mean fraction is $\sim$ 0.005, with a scatter
of a factor of 10. The red line in the same panel shows the results from
\citet{MaoJing04apj}, which were obtained from 12 haloes (of galactic, group
and cluster masses) and 30 random projections. Their result lies somewhat
higher than found here although still within the scatter. This is probably due
to the inclusion of group- and cluster-sized haloes in the averaging, which
tend to have a higher substructure fraction due to their later formation times.
The blue point indicates the required substructure mass fraction found
by \citet{DK2002} to be 0.02 (median, ranging from 0.006 to 0.07 at 90\% 
confidence).

\subsection{Adding ``light'' to dark matter haloes \label{sec:insertBaryon}}

We put the source redshift $z_{\rm s}$ at 3.0. This is reasonable since many
lensed quasars are at similar redshift. The lensing critical surface density
is given by
\begin{equation}
\CritDens = \frac{c^2}{4\pi G} \frac{\Ds}{\Dl\Dls},
\end{equation}
where $\Ds$, $\Dl$, and $\Dls$ are the angular diameter distances between the
source and the observer, the lens and the observer, and the source and the
lens, respectively. For our adopted source and lens redshifts, $\CritDens =
1.82 \times 10^9$ $M_{\odot}$ kpc$^{-2} = 7.95 \times 10^{10}$ $M_{\odot}$
arcsec$^{-2}$.

\begin{figure*}
\centering
\includegraphics[scale = 0.36]{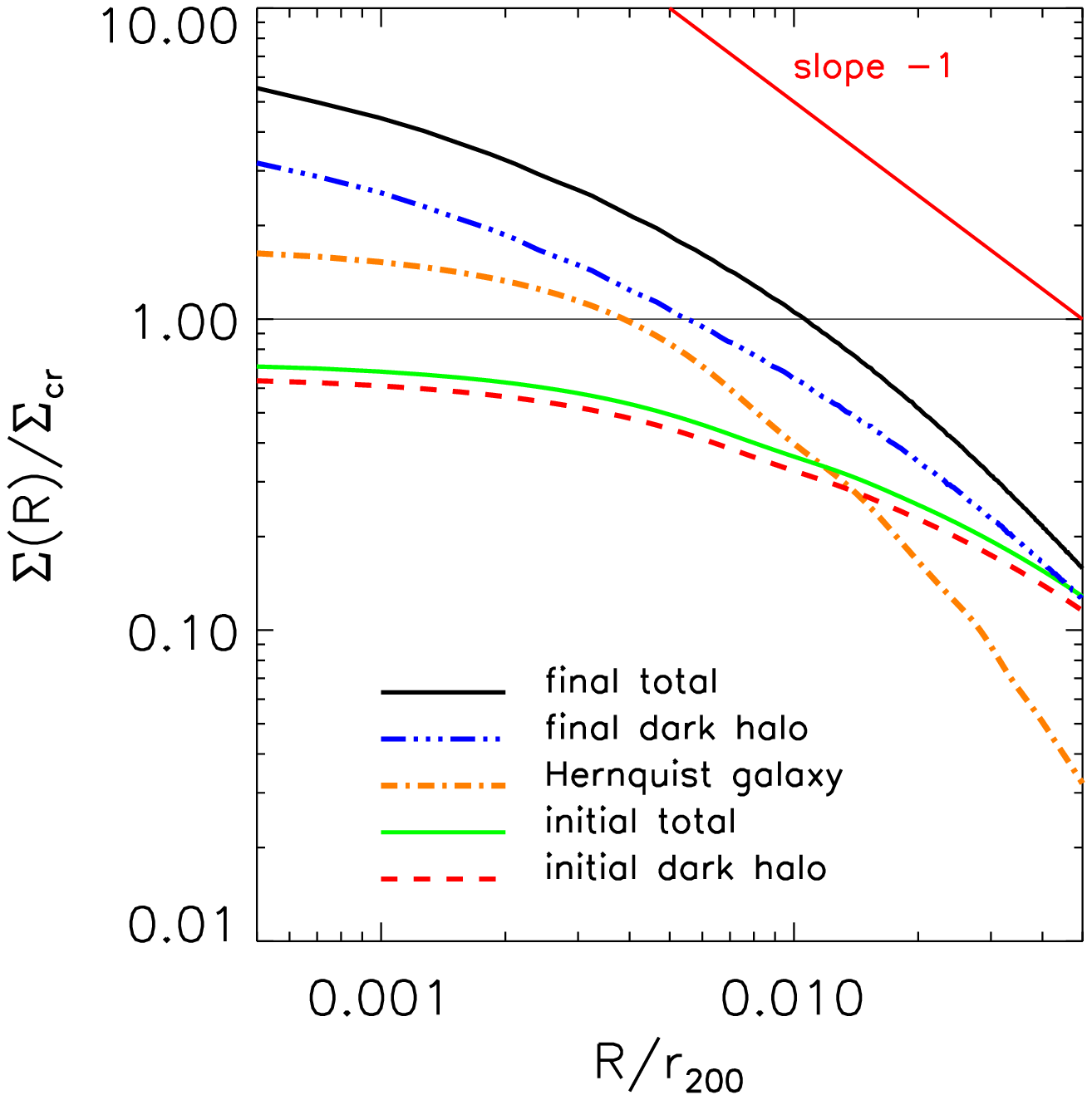}
\includegraphics[scale = 0.36]{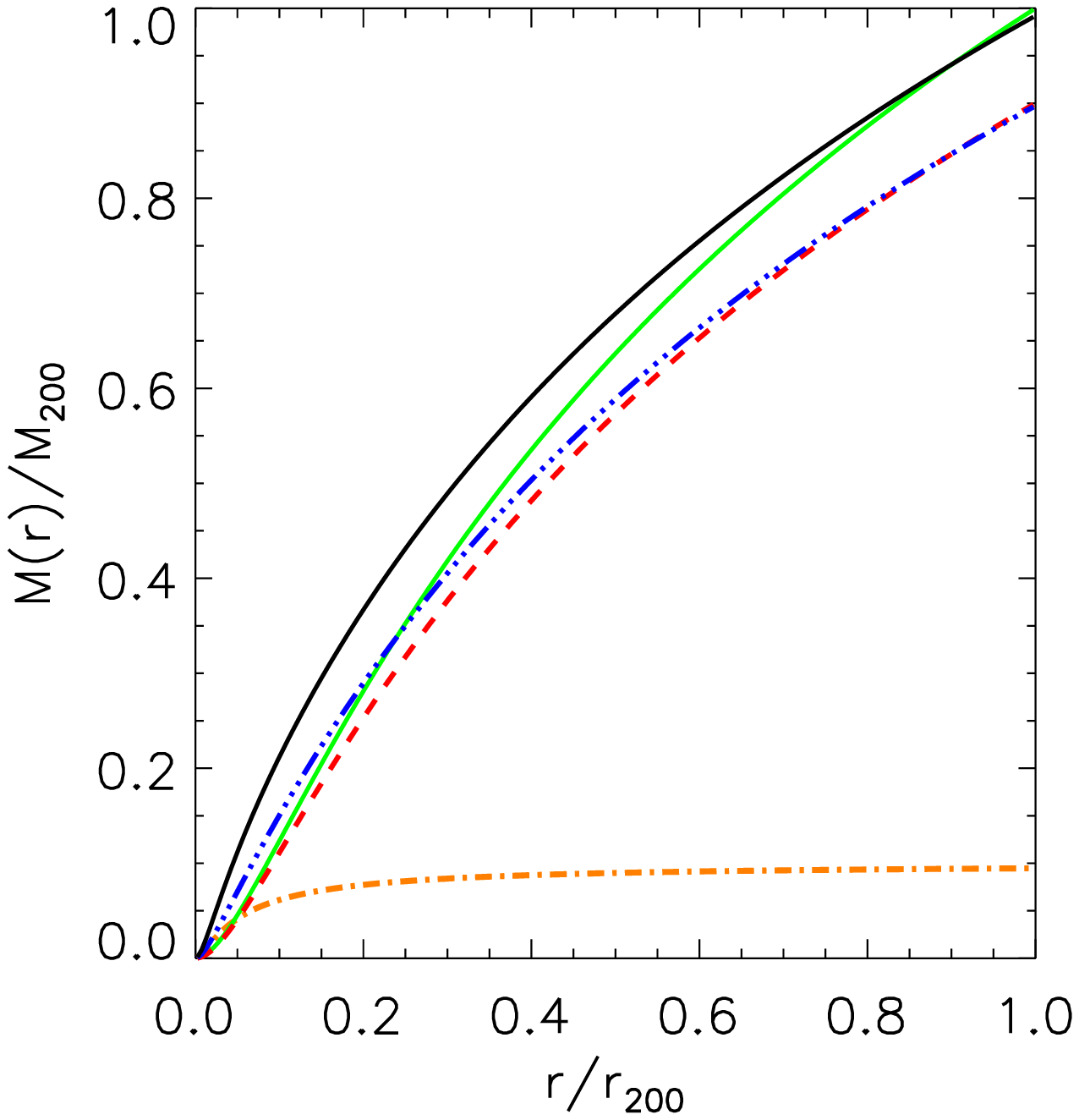}
\includegraphics[scale = 0.36]{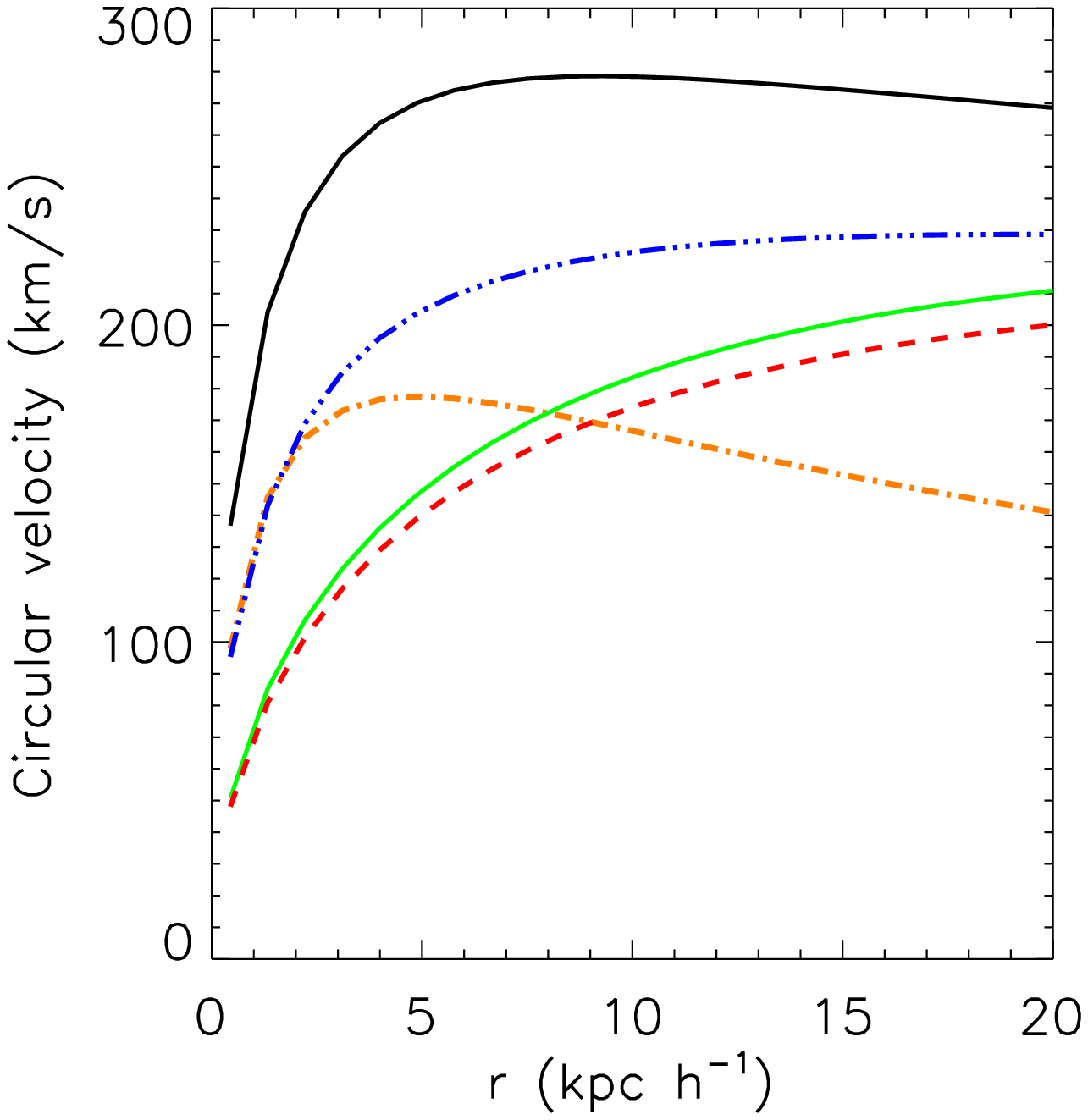}
\caption{The halo {\it Aq-D-2}: the left panel shows the surface density
  profiles $\Sigma(R)$ projected in the $Y$-direction, and normalised to the
  critical surface density.  Profiles are for cases before and after adding a
  Hernquist galaxy and the dark matter halo's adiabatic contraction, assuming
  the added baryonic component has 10\% of the total mass and an effective
  radius of 5\% of the halo virial radius ($\fcool=0.1$, $\fR=0.05$). Line
  symbols are labelled inside the figure. The isothermal slope ($\Sigma(R)
  \propto R^{-1}$) is indicated by the red line at the top right (see
  \S\ref{sec:insertBaryon} for details).  The middle panel shows the mass
  distributions $M(\leq r)$.  The right panel shows the rotation curves
  $V_{\rm c}(r)$. The final total rotation curve is flat from $\sim
  5h^{-1}\kpc$ out to a few tens of kpc.} \label{fig:Mainhalo2}
\end{figure*}

To produce multiple images, the maximum surface density of a halo usually has
to be super-critical. The left panel of Fig.~\ref{fig:Mainhalo2} shows the
surface density distribution for the halo {\it Aq-D-2} projected along the
$Y$-axis. Clearly, the central surface density of the (initial) NFW
dark matter halo is below the critical value, and thus generally
no multiple images can be produced (e.g. \citealt{Williams99ApJ,RusinMa2001}) .
This is hardly surprising, since for galaxy-scale strong lensing,
the images form only a few kpc (projected) from the centre where baryons play
a crucial role. Thus, one must incorporate a baryonic component in order to
model the lensing galaxies more realistically, a topic we turn to next.

Most gravitational lenses are early-type (elliptical) galaxies rather than
late-type (disk) galaxies, as the former are more massive and dominate the
lensing cross-sections (\citealt{TOG}).
There have been many hybrid models used for the lensing galaxies
(e.g. \citealt{Keeton2001, KochanekWhite2001, Oguri2002,
  JiangKochanek2007}). We use the spherical Hernquist profile to model
the light distribution, since it approximates the de Vaucouleur's profile
that has been observed for elliptical galaxies and bulges, and it has
many known, convenient analytical properties.

The three-dimensional density and mass profiles $\rho_H(r)$, $M_H(r)$ for the
Hernquist distribution are given by (\citealt{Hernquist1990apj}):
\begin{equation}
\begin{array} {c}
\displaystyle \rho_H(r) = \frac{a M_\star}{2\pi r} \frac{1}{(r+a)^3}, \\
\displaystyle M_H(<r) = M_\star \frac{r^2}{(r+a)^2}.
\end{array}
\label{eq: HernquistEquation}
\end{equation}
where $M_\star$ is the total baryonic mass, and $a$ is a scale length
related to the effective spherical radius $\Reff$ (within which half
of the mass is contained) by $a=\Reff/(\sqrt2+1)$.

The profile is specified by two parameters $a$ (or $\Reff$) and $M_\star$,
which are linked with the dark matter halo
parameters $r_{200}$ and $M_{200}$ by
\begin{equation}
  \begin{array} {c}
    \displaystyle \fR = \frac{\Reff}{r_{200}}, ~~
    \displaystyle \fcool = \frac{M_\star}{M_{200}}, ~~
    \displaystyle M_{200} = M_{\rm DM} + M_\star.
  \end{array}
  \label {eq:By2DmEquation}
\end{equation}
Notice that the mass of the main halo dark matter $M_{\rm DM}$ is reduced
by a factor of (1-$\fcool$) to conserve the total mass and the
mass of substructures.

The inclusion of the baryonic component affects the distribution of the dark
matter halo. Many studies have shown that the adjustment of the dark matter
halo can be approximated by an adiabatic contraction
(\citealt{BarnesWhite1984}; \citealt{Blumenthal1986AC}).
\citet{Gnedin2004} have proposed a modification to this simple model
in order to take into account the fact that particle orbits in realistic 
halos are not circular, but it is not clear whether this modification 
is able to reproduce accurately the results of numerical simulations 
(see, e.g. \citealt{Abadi2009}). In view of this, we have decided to 
follow, for simplicity, the procedure outlined by \citet{MoMaoWhite1998}.
Assuming that both the baryon and dark matter components follow an NFW 
distribution initially, baryons ($\fcool$ percent of the total matter) 
then cool to form the galaxy at the centre, which causes the dark 
matter halo to contract adiabatically. After the adiabatic contraction,
the dark halo follows a new profile and hosts a Hernquist galaxy at its
centre. Note that we contract all the particles in different
components (i.e., diffuse dark matter and subhaloes) in the same way.

The two parameters ($\fR$ and $\fcool$) are chosen according to two
criteria (after adding the baryonic galaxy and accounting for the
adiabatic contraction): (1) the projected dark-matter mass fractions
inside the Einstein radii of the host galaxy haloes should range
from 0.4 -- 0.7 (\citealt{TreuKoopmans04apj}); (2) the projected
slopes are close to isothermal at a few kpc from the galactic centre
(e.g. \citealt{Rusin2003, RusinKochanek2005, Koopmans2006apj,
Gavazzi2007}), or equivalently, the final rotation curves are
roughly flat from a few kpc out to a few tens of kpc (see
Fig.~\ref{fig:Mainhalo2}). Furthermore, $\fcool$ should be smaller
than the universal baryonic fraction of $\sim$ 17.5 per cent (from
WMAP-3, \citealt{Spergel2007WMAP3}).

We find that $\fR=0.05$ and $\fcool=0.1$ satisfy these criteria well.  From
Fig.~\ref{fig:Mainhalo2} (the left panel), it is clearly seen that after
inserting the baryonic galaxy and taking the adiabatic contraction into
account, the total surface density is now super-critical and the
corresponding Einstein radius is of the order of a few kpc, similar to that in
many gravitational lenses. Notice however that our procedure is not
self-consistent dynamically, since the inclusion of a baryonic component will
affect the evolution and survival of subhaloes. We shall return to this point
briefly in the discussion.

\section{LENSING METHODOLOGY}

$N$-body simulations provide us with the positions (and velocities) of
particles. For lensing calculations, we first project the particles onto a
mesh in the lens plane (and tabulate the stellar surface density,
and then smooth the surface density
field appropriately. Using the smoothed surface density map,
we can numerically calculate the lensing potential,
deflection angles and magnifications. The details of the numerical
procedure are given in \S\ref{sec:method}.

We test the accuracy of our numerical procedure by comparing with known
analytical results, using a singular isothermal sphere realised through Monte
Carlo simulations in \S\ref{sec:SIS}. We then relax the spherical assumption,
and further test our procedure with an isothermal ellipsoid generated with a
similar number of particles as those in the Aquarius simulations; the
comparison results are presented in \S\ref{sec:SIE}.

\subsection{From particles to lensing images \label{sec:method}}

\subsubsection{Coarse and fine particle meshes \label{sec:mesh}}

We use a Particle-Mesh (PM) code for the lensing potential calculation. The
application of Fast Fourier Transforms ($FFT$) in the PM algorithm makes it
computationally efficient. However it is limited in resolution by the finite
mesh size and so cannot accurately represent regions with rapid density
variations on the scale of the grid size. To increase the accuracy in the
regions of interest (within a few kpc from the centres of galaxies), we
establish two two-dimensional [2D] meshes: a coarse grid used for the
potential field generated by the mass projected outside the central ($20
h^{-1}$ kpc)$^2$ region, and a fine grid for the mass within. 
Both grids have $1024 \times 1024$ pixels, 
covering $(4 \, r_{200})^2$ and ($40 h^{-1}$ kpc)$^2$ (see \S 3.1.3) 
with resolutions $\sim$ 0.6$h^{-1}$ kpc and 0.04$h^{-1}$ kpc for the coarse 
and fine grids, respectively (the factor of two increase in the box size 
is due to the isolated boundary condition, see \S\ref{sec:isolated}).
This resolution ensures that the tangential critical curves are resolved with
sufficient accuracy. In contrast, the inner radial critical curves may not be
well reproduced, due to the finite resolution of the mesh. However, this is
not a major concern since all the bright images that we are interested in form
close to the outer (tangential) critical curves. Furthermore, the resolution
of the fine mesh is similar to the softening length of the simulations, and
the density distributions in the very central regions are not accurately
modelled in the simulations on smaller scales than the gravitational softening
in the first place.

\subsubsection{Particle assignment with smoothed particle hydrodynamics kernel}

The surface density maps of the Aquarius haloes are obtained by
assigning particles to the potential meshes using the Smooth Particle
Hydrodynamics (SPH) kernel (\citealt{Monaghan1992SPH}).  Although, in the
end, we will approximate the underlying mass distributions of the Aquarius
haloes by isothermal ellipsoids in order to circumvent problems caused by
discreteness noise (see \S\ref{sec:SIE}), SPH-smoothed density fields are
used as an intermediate step to generate basic lensing properties (e.g.
critical curves and caustics) to constrain the best-fit isothermal
ellipsoids. For more detail, see \S\ref{sec:aquarius}.

The advantage of the SPH assignment is that it adjusts the smoothing scale
according to the local density environment: particles in a high density region
are mildly smoothed while those in a low density region are smoothed more.
For each particle, a smoothing length $\hsml$ is calculated
according to the local number density in its 3D neighbourhood. The
particle mass is then assigned to all the mesh cells that are within
a circle of $2\hsml$ in radius in its neighbourhood. The 3D density
kernel can be integrated along the line of sight analytically to
obtain the surface density distribution:
\begin{equation}
\Sigma(u) = \frac{1}{\pi \hsml^2} \left\{{\begin{array}{*{20}l}
\frac{1}{16}[-(8+52u^2)\sqrt{1-u^2}+(16+26u^2)\sqrt{4-u^2}\\
-9u^4\ln u+3u^2(16+4u^2)\ln(1+\sqrt{1-u^2}) \\
-3u^2(16+u^2)\ln(2+\sqrt{4-u^2})], ~~ {\mathrm{if}~ 1 > u \geq 0} \\
\frac{1}{16}[2\sqrt{4-u^2}(8+13u^2)+3u^2(16+u^2)\ln u \\
-3u^2(16+u^2)\ln(2+\sqrt{4-u^2})], ~~ {\mathrm{if}~ 2 > u \geq 1}\\
0, ~~~ {\mathrm{if}~ u > 2}
\end{array}}\right.
\end{equation}
where $u \equiv r/\hsml$ is the distance from the cell centre to the particle
normalised to $\hsml$, and the total mass within $u \leq 2$ is unity.

The smoothing length $\hsml$ for each particle depends on its local density
and is controlled by the parameter $N_{\rm ngb}$, the number of particles that
are contained within radius $\hsml$. A good smoothing procedure
should reduce the numerical noise without smoothing excessively out the real
density fluctuations (e.g. substructures).

The total mass that has been assigned to the neighbouring cells should be
equal to the mass of the particle. However this is only approximately true due
to the discreteness of cells. In particular, mass conservation is quite poorly
observed when the smoothing length $\hsml$ is only a few mesh cells, which may
happen in a dense environment. For particles with $2 \hsml \le 15$ cell sizes
(30 cells in diameter), we therefore renormalise each individual kernel so
that the total mass is conserved during the assignment.

We find in practice that SPH assignment is superior to Cloud-In-Cell
(CIC) assignment in terms of reducing discreteness noise. For a singular
isothermal sphere realised with $10^6$ particles, the SPH-smoothed ($N_{\rm
  ngb}=32$) and CIC-smoothed surface density fields show fluctuations of 2\%
and 30\% relative to the analytical results, respectively. For a realisation
with $10^7$ particles, the fluctuations decrease to 1\%  for the SPH
assignment ($N_{\rm ngb}=320$, with the same smoothing length, $\hsml
\propto N_{\rm p}^{-1/3} N_{\rm ngb}^{1/3}$ (\citealt{Li2006SPH})) and
to 10\% for the CIC assignment.

\subsubsection{Isolated boundary conditions \label{sec:isolated}}

Periodic boundary conditions are most natural for Fourier Transforms, but are
not appropriate for lensing galaxies.  We follow
\citet{HockneyEastwood1981book} to eliminate the (aliasing) effects due to
``mirror'' particles by using a mesh twice as big as the simulated lens
system, padding the region outside the simulation volume with zeros. A
truncated 2D gravitational force kernel is tabulated onto the same simulation
mesh, and then convolved with the assigned surface density field.  The
gravitational effect is accurately reproduced within
the region where the mass has been distributed (See
\citealt{HockneyEastwood1981book}, for more technical details).
We adopt this procedure throughout this work.

\subsubsection{Lensing potential, deflection angle and magnification}

After the discretisation of the surface density field through SPH assignment
and the tabulation of the truncated 2D gravitational kernels on the meshes,
the potentials and their derivatives are easily calculated by convolutions
which can be efficiently implemented in Fourier space.

In particular, the effective lensing potential $\psi(\vec{\theta})$ is the
convolution of the surface density $\Sigma(\vec{\theta})$ and the 2D kernel
$\ln |\vec{\theta}|$:
\begin{equation}
\psi(\vec{\theta}) = \frac{1}{\pi} \int \Sigma(\vec{\theta^{\prime}})\ln
|\vec{\theta}-\vec{\theta^{\prime}}|\,d^2\theta^{\prime}.
\end{equation}
The deflection angle $\vec{\alpha}(\vec{\theta})$ is the first
derivative of the lensing potential, $\psi(\vec{\theta})$, and is thus
the convolution of the surface density $\Sigma(\vec{\theta})$ and the 2D
force kernel $\vec{\theta}/|\vec{\theta}|^2$:
\begin{equation}
\vec{\alpha}(\vec{\theta}) \equiv \nabla \psi(\vec{\theta})=
\frac{1}{\pi} \int
\Sigma(\vec{\theta^{\prime}})\frac{\vec{\theta}-\vec{\theta^{\prime}}}
{|\vec{\theta}-\vec{\theta^{\prime}}|^2}\,d^2\theta^{\prime}.
\end{equation}
The convergence $\kappa(\vec{\theta})$ (the surface
density normalised to $\CritDens$) and the shear $\gamma(\vec{\theta})$
are second-order derivatives of the lensing potential
$\psi(\vec{\theta})$:
\begin{equation}
\begin{array} {c}
\displaystyle \kappa = (\psi_{11} + \psi_{22})/2,~~
\displaystyle \gamma_1 = (\psi_{11} - \psi_{22})/2, \\
\displaystyle \gamma_2 = \psi_{12} = \psi_{21}, ~~
\displaystyle \gamma^2 = \gamma_1^2 + \gamma_2^2,~~
\displaystyle ~ \psi_{ij}\equiv
\frac{\partial^2\psi}{\partial\theta_i\partial\theta_j},
\end{array}
\end{equation}
where the derivatives are taken with respective to the index 1 ($x$)
and 2 ($y$). Numerically, the convergence and shear can be calculated
through 4th-order finite differencing from the deflection angle
$\vec{\alpha}(\vec{\theta})$. The magnification $\mu(\vec{\theta})$
is related to the convergence and shear by
\begin{equation}
\mu = \frac{1}{(1-\kappa)^2-\gamma^2}.
\end{equation}

\subsubsection{Image finding and cusp relation}

Since all the lensing quantities are now known, it is straightforward to find
the images for any given source position. To this end, we construct a separate
mesh in the image plane, with a resolution ($0.02 h^{-1}$ kpc) higher than the
fine potential mesh discussed in \S\ref{sec:mesh}; the lensing properties
(deflection angle, magnification etc.) on this ultra-fine mesh are found
through bi-linear interpolation. We then search image positions (and
magnifications) using the Newton-Raphson and triangulation methods
(\citealt{SchneiderBook1992}).

Of particular interest to gravitational lensing are the critical curves and
caustics. Critical curves in the image plane are a set of points where the
magnification is formally infinite for a point source, $\mu(\vec{\theta})
\longrightarrow\infty$. In practice, they are identified according to the fact
that the magnifications have different signs (i.e., different parities) for
images on different sides of a critical curve. Critical curves are mapped into
caustics in the source plane, which can be easily obtained through the lens
equation.

Most strong lenses occur in elliptical galaxies since they have larger
lensing cross-sections than spiral galaxies (\citealt{TOG}). They
typically form two distinct sets of critical curves and corresponding
caustics: the tangential (``outer'') and radial (``inner'') critical
curves, which are mapped into tangential (``inner'') and radial
(``outer'') caustics (see Fig.~\ref{fig:MCIE_Rcusp1} for an
example). A source inside the central caustic usually produces five
images: four close to the tangential critical curve and one central
image which is usually too faint to be observable (and is of no
interest to us for the present work).

We are particularly interested in sources that are close to the cusps of the
tangential caustic (``cusp sources''). For cusp sources, three close images
form around the tangential critical line, with alternate parities. There are
two different kinds of cusp sources and corresponding image configurations. As
illustrated in Fig.~\ref{fig:MCIE_Rcusp1}, a ``major cusp'' source forms three
images around the tangential critical curve on the same side of the source
(with respect to the centre of the lens) while a ``minor cusp'' source forms
three close images on the opposite side of the source.

In any smooth lensing potential, for a source very close to a cusp, the three
close images satisfy an asymptotic magnification relation (the ``cusp-caustic
relation''; \citealt{BN1986apj}; \citealt{SW1992aa}; \citealt{KGP2003apj}):
\begin{equation}
  \Rcusp \equiv \frac{|\mu_A + \mu_B +
    \mu_C|}{|\mu_A|+|\mu_B|+|\mu_C|} \rightarrow 0,
\end{equation}
with the total absolute magnification $|\mu_A|+|\mu_B|+|\mu_C|
\rightarrow \infty$.

For each of the cusp sources, we define an image opening angle $\Delta\theta$,
ranging from 0 to $\pi$, which measures the angle (from the lens centre) of
the outer images of the close triple.  Notice that both $\Delta\theta$ and
$\Rcusp$ are observable.  In a smooth lens potential, as a source moves to a
cusp caustic, both $\Delta\theta$ and $\Rcusp$ decrease asymptotically to
zero.  As can be seen from Fig.~\ref{fig:MCIE_Rcusp2}, there are two leading
patterns on the $\Rcusp-\Delta\theta$ diagram due to ``major'' and ``minor''
cusp sources. Generally speaking, the major cusp sources have larger $\Rcusp$
than the minor cusp sources for the same image opening angle.

The cusp-caustic relation predicts that in smooth lens models $\Rcusp$ would
asymptotically approach zero when a source moves towards the caustic. However,
the presence of (clumpy) substructures will break down the smooth potential
assumption in the asymptotic cusp-caustic relation, resulting in substantial
deviations in $\Rcusp$ values and other quantities (such as image positions
and time delays) from simple predictions. Therefore the examination of the
cusp-caustic relation is a way to test for the presence of substructures that
are projected near the (tangential) critical curves. However, caution must be
exercised because, even for smooth lens models, a high $\Rcusp$ is possible.
There are many factors that affect the $\Rcusp$ distribution apart from the
presence of substructures, e.g. the mass distribution of the lens (radial
profile and the ellipticity), external shear from the environment, and the
selection criteria of the cusp sources (for more discussions see
\citealt{KGP2003apj}).

\subsection {Singular isothermal sphere \label{sec:SIS}}

\begin{figure*}
\centering
\includegraphics[scale = 0.33]{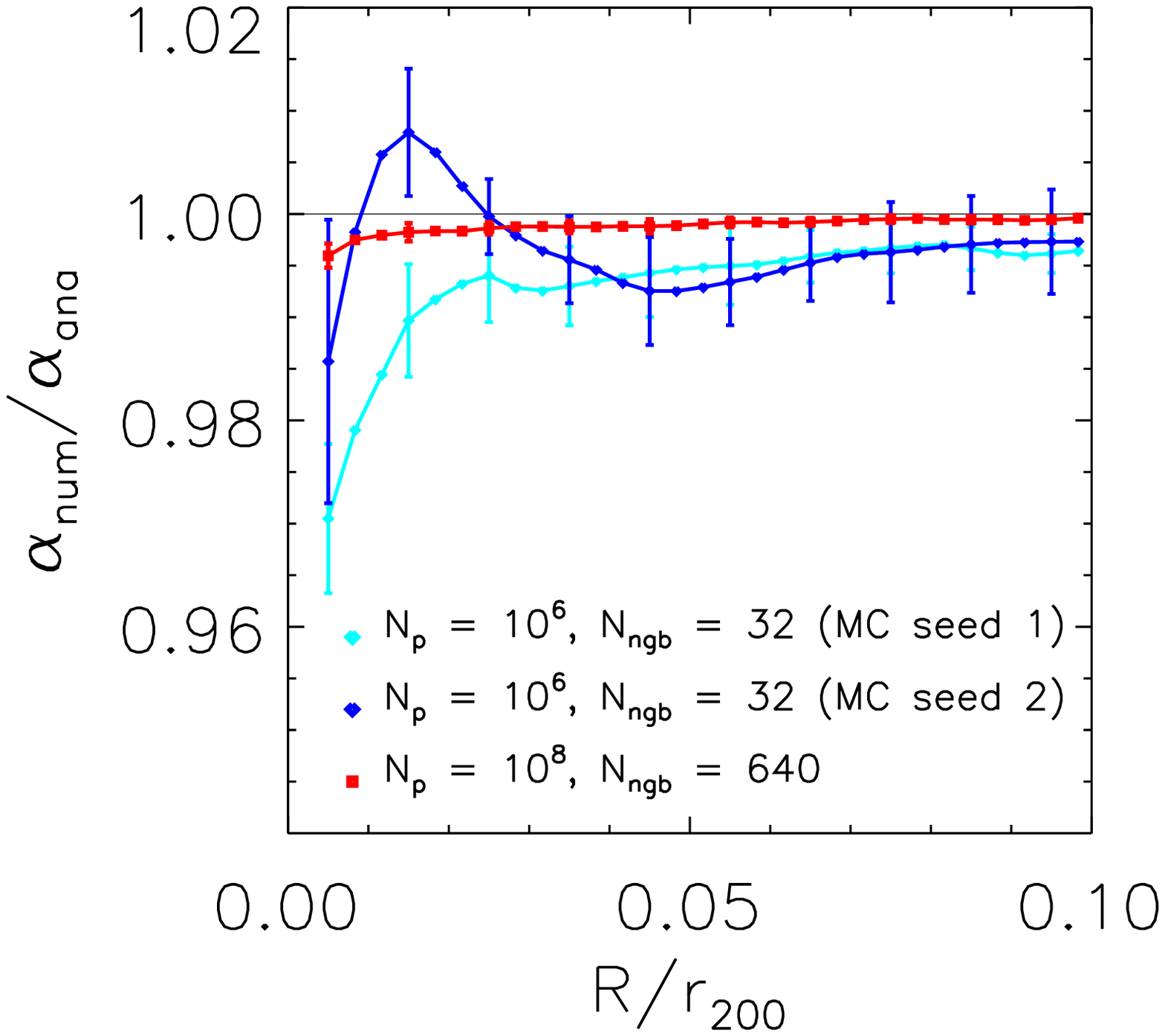}
\includegraphics[scale = 0.33]{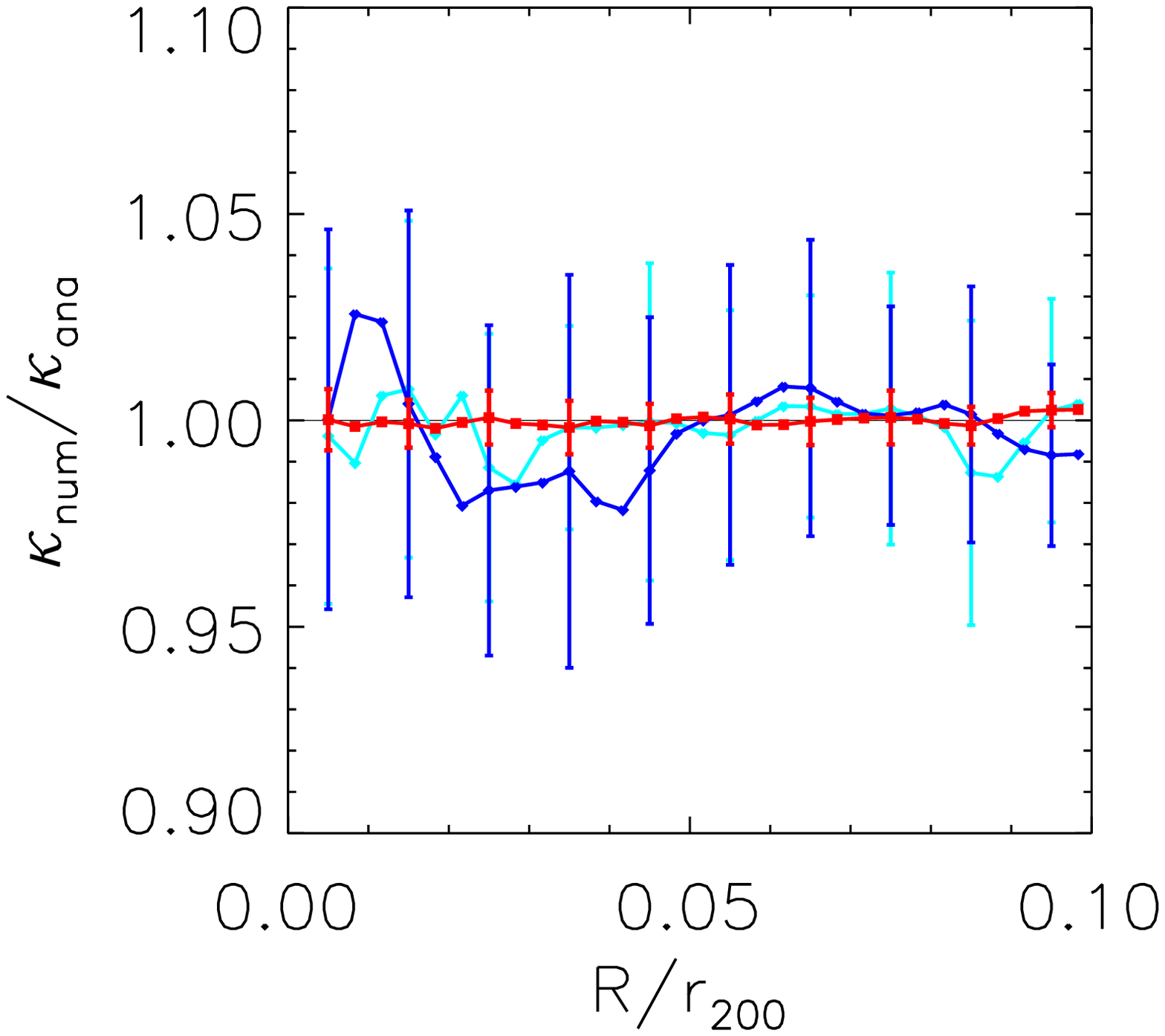}
\includegraphics[scale = 0.33]{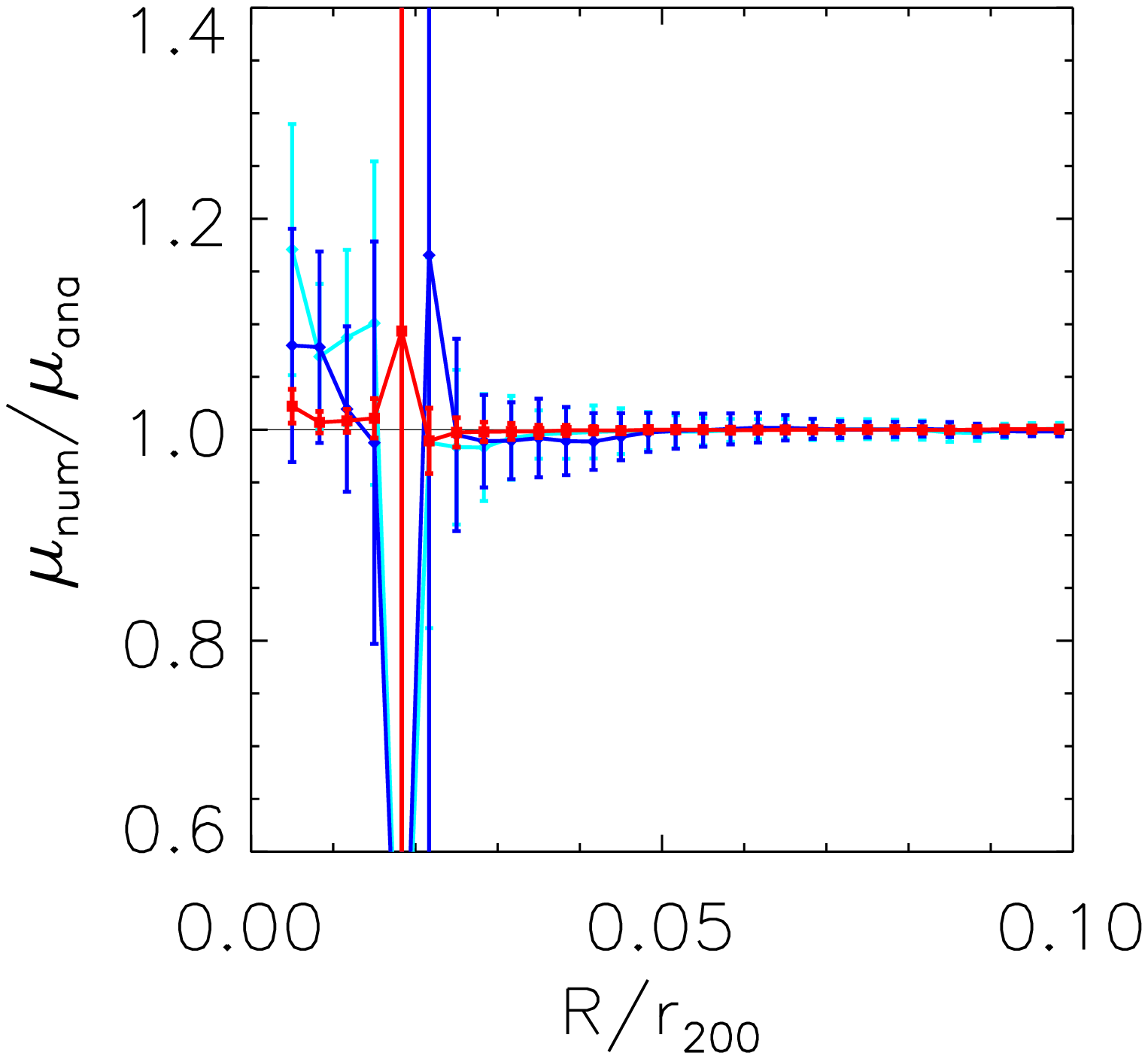} \\
\includegraphics[scale = 0.33]{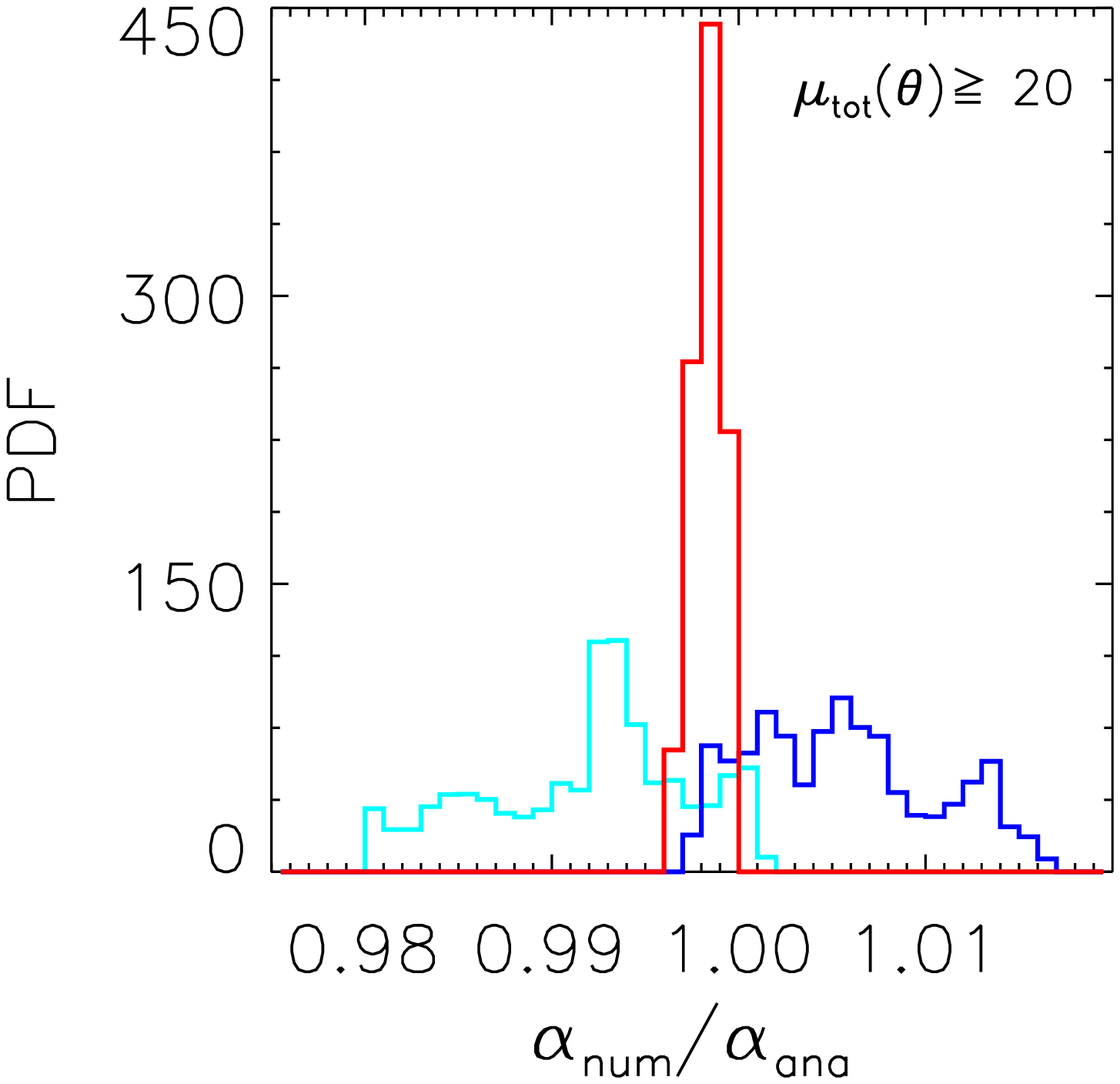}
\includegraphics[scale = 0.33]{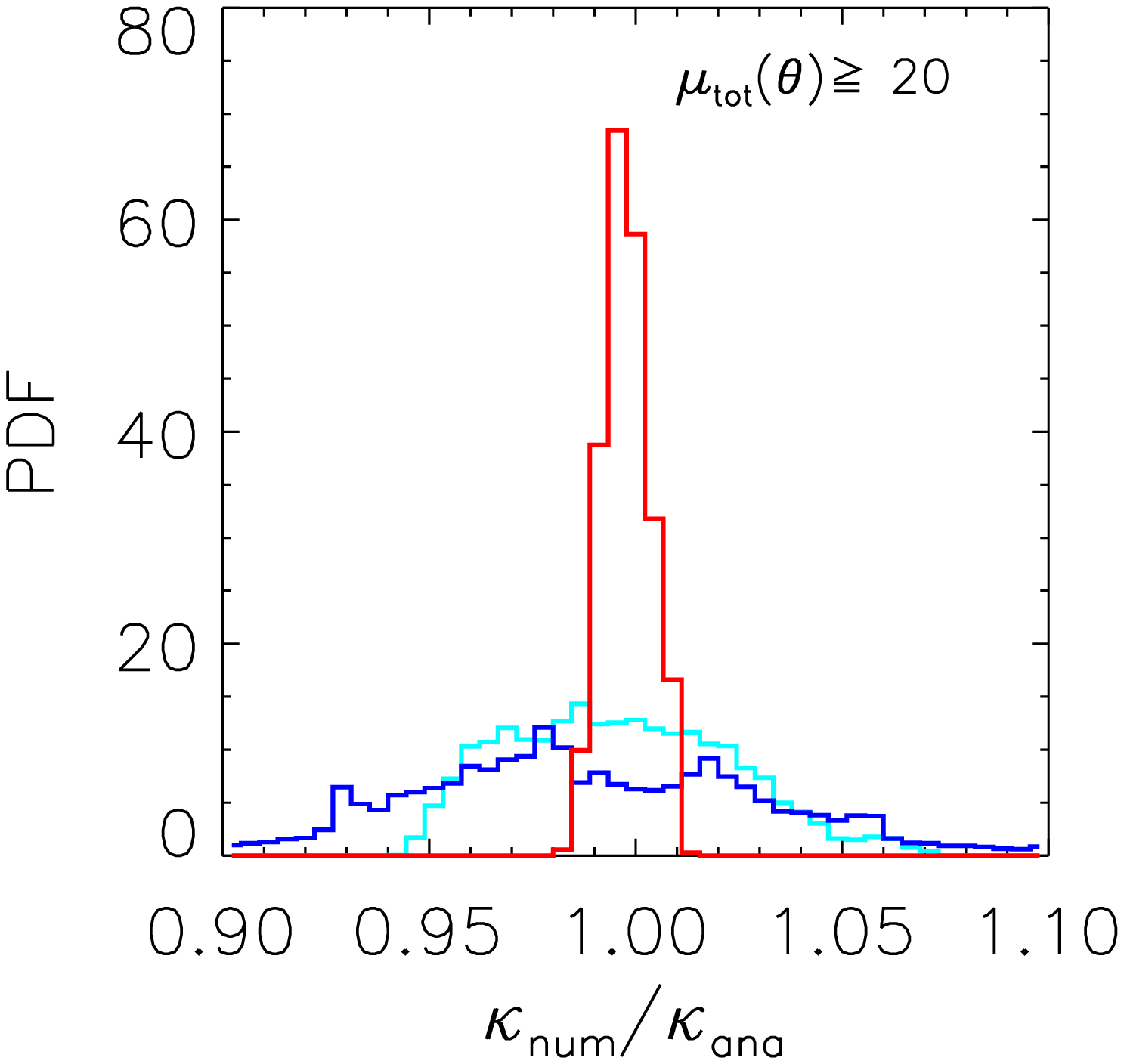}
\includegraphics[scale = 0.33]{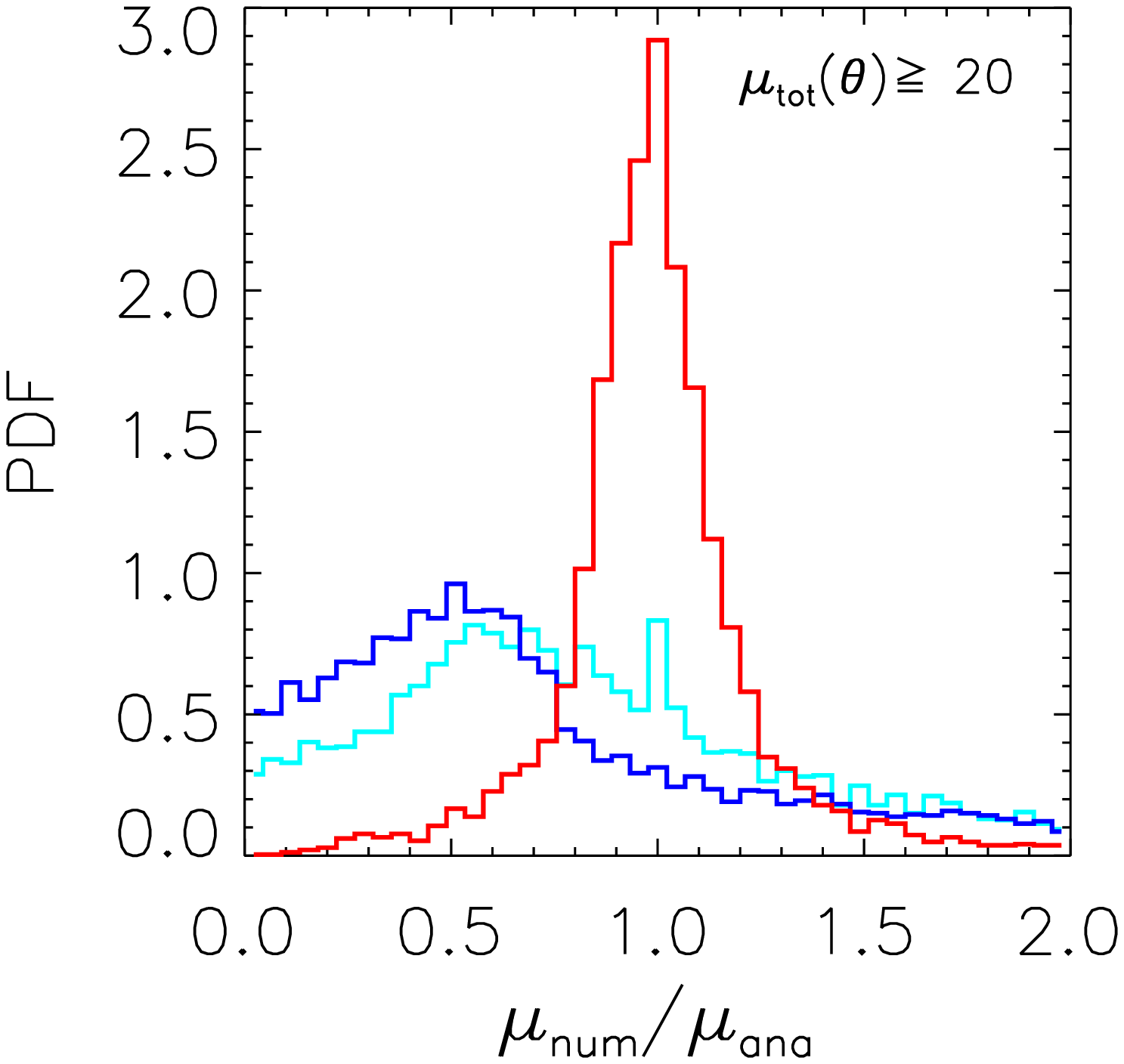}
\caption{The numerical accuracy of the deflection angle, the convergence, and
  the magnification for Monte-Carlo realisations of singular isothermal
  spheres. The top panels show the ratios of the numerical
  to the analytical results as a
  function of radius. The deviation in the numerical magnification (on the
  right) towards the centre is due to the finite mesh resolution of the
  Particle-Mesh code, and that seen near the Einstein radius (at about 0.02
  $r_{200}$) is due to the divergent behaviour of the magnification close to
  the critical curve. The corresponding probability distributions for images
  with a total magnification above 20 (around $r \sim 0.02$ $r_{200}$) are
  presented in the bottom panels. The cyan and blue curves are for two
  $10^6$-particle realisations (with $N_{\rm ngb}=32$) while the red curve is
  for a $10^8$-particle realisation (with $N_{\rm ngb}=640$).
} \label{fig:Test_SIS}
\end{figure*}

We test our lensing simulation code with Monte-Carlo realisations of  singular
isothermal spheres (SIS), for which analytical lensing properties are
known. Each of our SIS contains a mass of $10^{12} h^{-1}M_{\odot}$ within a
virial radius of 100$h^{-1}$ kpc, realised with $10^6$ and $10^8$ particles;
the SPH assignment parameter is chosen to be $N_{\rm ngb} = 32$ and $N_{\rm
  ngb} = 640$ for the two cases respectively. Fig.~\ref{fig:Test_SIS} shows
the numerical accuracy of the deflection angle, convergence (surface density)
and magnification in our numerical procedures. For the $10^6$ particle case,
two Monte-Carlo realisations (cyan and blue curves) are shown.  For the $10^8$
particle realisation, the uncertainties around the Einstein radius (at about
0.02 $r_{200}$, defined by a total magnification $\mu(\vec{\theta}) \geq 20$)
are 0.2\% for the deflection angle, 1\% for the convergence, and
$<$ 10\% for the lensing magnification (estimated by the half width
half maximum of the probability distributions). The deviation towards the
centre is due to the fact that the finite mesh resolution of the Particle-Mesh
code fails to represent the singular behaviour at the centre of the SIS. The
significant deviation of the magnification seen near the Einstein radius is
due to the divergent behaviour of the magnification close to the critical
curve, where $\mu = 1/((1-\kappa)^2-\gamma^2) \longrightarrow \infty$, when
$\kappa = \gamma = 0.5$ at the Einstein radius for the singular isothermal
sphere.

\subsection{High-resolution isothermal ellipsoid \label{sec:SIE}}

We simulate an isothermal ellipsoid (IE) with $10^6$ and $10^8$
particles (as in the Aquarius haloes).
Such an isothermal ellipsoidal distribution is modelled
as an oblate spheroid with axis ratio $q_3$ and with a density distribution:
\begin{equation}
  \rho \propto (S_0^2+R^2+z^2/q_3^2)^{-1},
  \label {eq:aboutrho}
\end{equation}
where $S_0$ is a core radius, and
$(R, z)$ are the cylindrical coordinates. It is specified by three parameters
(see \citealt{KK98IE} for details): the effective critical radius $b_{\rm I}$,
the eccentricity of the mass distribution $e = (1-q_3^2)^{1/2}$, and a core
radius $S_0$. The parameters for the isothermal ellipsoidal halo are adjusted
so that its critical curves and caustics match those for the halo {\it Aq-F-2}
in the $z$-projection.  The parameters are $b_{\rm I} = 0.4\arcsec$, $q_3 =
0.8$, $S_0 = 0.1\arcsec$ and the major-axis of the surface density ellipse is
rotated by $\sim \pi/8$ with respect to the $X$-axis.

\begin{figure*}
\centering
\includegraphics[scale = 0.75]{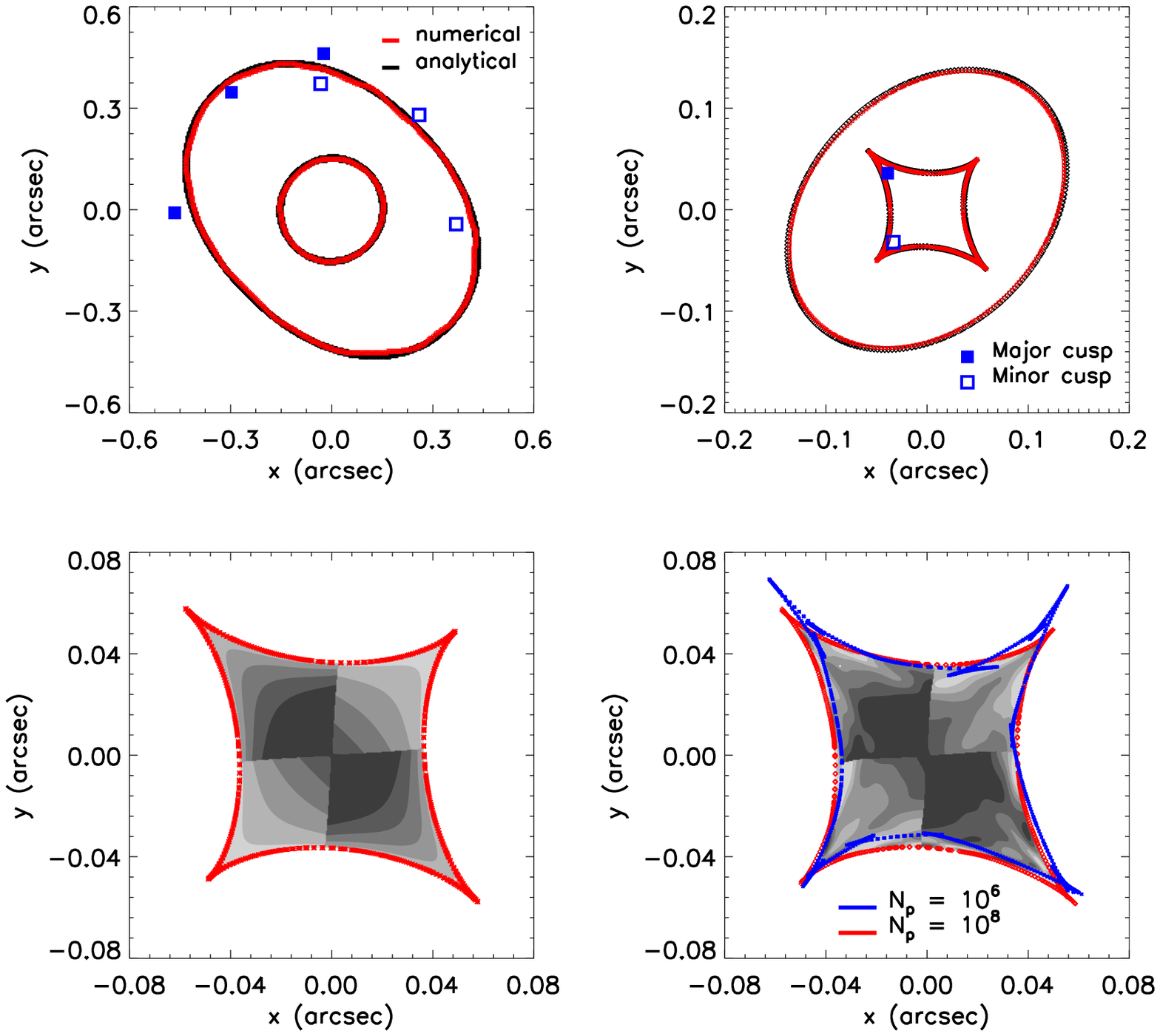}
\caption{Critical curves, caustics, and cusp-caustic relation $\Rcusp$ maps
  for a Monte-Carlo realisations of isothermal ellipsoid
  with $N_{\rm p} = 1.6\times 10^8$ and $N_{\rm ngb}=640$.
  The top panels show the critical curves and the
  caustics. The position and corresponding images are shown for a ``major
  cusp'' (solid squares) and a ``minor cusp'' source (open squares).
  The bottom panels show the $\Rcusp$ maps from the analytical
  solution (left) and from the numerical result (right), with contour
  levels (0.0, 0.05, 0.1, 0.15, 0.2).
  The numerical tangential caustic from a $N_{\rm p} = 10^6$
  Monte-Carlo realisation is also presented (blue curve). The
  swallow-tails due to numerical noise are more apparent in this
  case.} \label{fig:MCIE_Rcusp1}
\end{figure*}

\begin{figure*}
\centering\includegraphics[scale = 0.55]{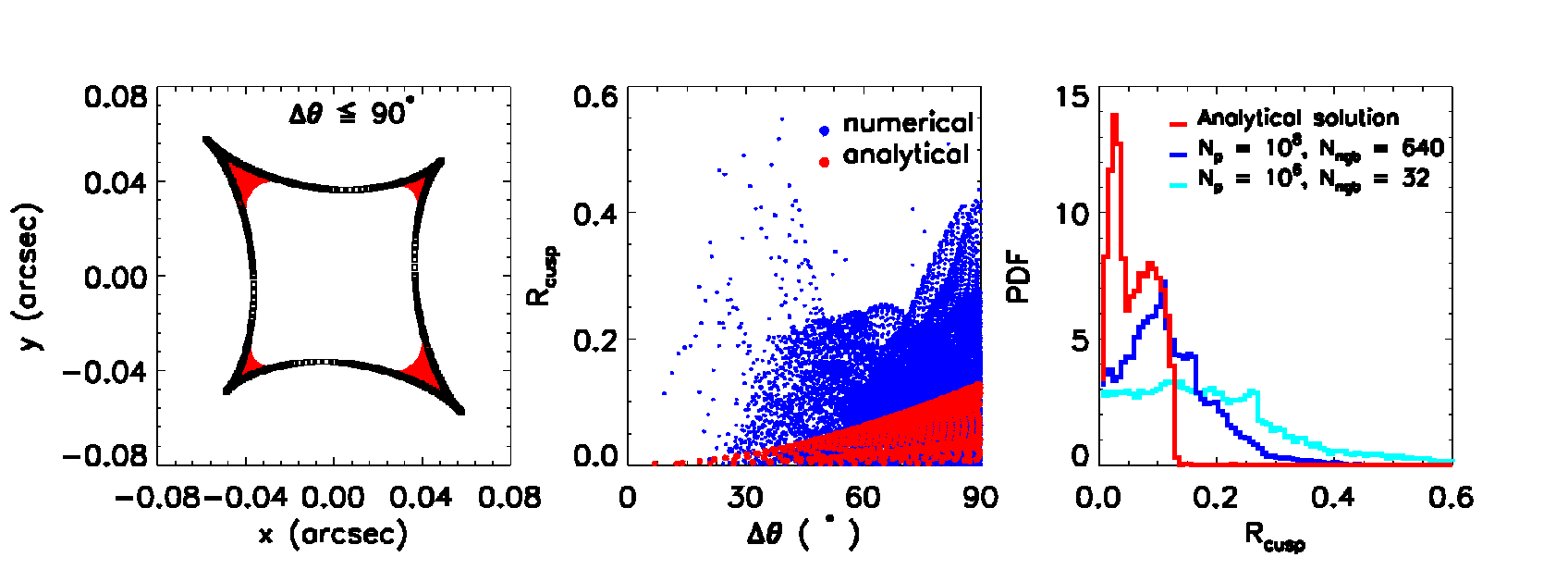}
\caption{The cusp-caustic relations (of
  the same Monte-Carlo realised isothermal ellipsoidal (IE) halo as in
  Fig.\ref{fig:MCIE_Rcusp1}) for caustic sources with image opening angles
  $\Delta\theta \leq 90 ^{\circ}$.
  The left panel shows the source positions with respect to the
  tangential caustic. The middle panel shows both the numerical
  ($10^8$-particle realisation) and analytical results of $\Rcusp$ vs.
  $\Delta\theta$. The right panel shows the probability density
  distributions of $\Rcusp$ for the analytical IE (red), and the two
  Monte-Carlo realised haloes with $10^6$ (cyan) and $10^8$ particles
  (blue), respectively.} \label{fig:MCIE_Rcusp2}
\end{figure*}

Fig.~\ref{fig:MCIE_Rcusp1} shows analytical and numerical critical curves,
caustics and $\Rcusp$ maps for sources located inside the diamond
caustics. The two critical curves nearly overlap each other, with barely
noticeable wiggles in the numerical result. The caustics also agree reasonably
well for the $10^8$-particle case, but for the $10^6$-particle realisation
(blue lines), higher-order singularities such as swallowtails are clearly seen
close to the cusps and along the fold caustics. These arise due to numerical
noise. The numerical $\Rcusp$ map shows visible distortions compared with the
smooth contours in the analytical results, even in the central region.

Fig.~\ref{fig:MCIE_Rcusp2} presents the $\Rcusp$ distributions for caustic
sources (indicated in the left panel) with image opening angle $\Delta\theta
\leq 90^{\circ}$. The analytical results show two distinct peaks due to
major-cusp and minor-cusp sources with a sharp dropoff around $\Rcusp \sim
0.12$. In contrast, for the numerical distributions, even the $10^8$-particle
realisation shows a much broader profile than the analytical one, with an
extended tail out to $\Rcusp\sim 0.4$ due to numerical noise.  Notice that the
numerical noise behaves in a similar way as real substructures in the $\Rcusp$
distribution.

In the current numerical set-up with $N_{\rm ngb} = 640$ for a $10^8$-particle
simulation, the smoothing length $\hsml$ in the central region (of the halo)
approximately reaches the softening length of the Aquarius simulation, also
roughly the cell resolution of the fine mesh. Increasing $\hsml$ will indeed
further suppress the noise, but it may also over-smooth the underlying density
field. Below we outline an alternative way to approximate the smooth underlying
density fields for the host galaxy haloes.

\section{Results} \label{sec:aquarius}
\subsection{Lensing Predictions for Aquarius Haloes}
In this section, we will apply our lensing methodology to the Aquarius haloes
(and a baryonic component modelled as a Hernquist profile), and study the
violation of the cusp-caustic relation due to the dark matter substructures
therein. In principle, we should compare the lensing properties of the
simulated haloes with and without substructures in order to assess the
effects of substructure. However, Fig. \ref{fig:MCIE_Rcusp2} shows a
substantial broadening of the $\Rcusp$ distribution due to numerical noise,
which will significantly confuse signals from real substructures.

As mentioned above, the total (dark matter plus baryons) mass profile
of each halo is adjusted to resemble an isothermal distribution in the
central region. To avoid excessive discreteness noise, we go one step further
and adopt a fitted isothermal ellipsoid (with a density distribution given 
by eq. [\ref{eq:aboutrho}]) rather than the original particle
distribution for subsequent lensing calculations.
The parameters of the isothermal ellipsoid model are adjusted to match the
original critical curves and the caustics of each Aquarius halo (together
with the central galaxy) in a particular projection. We add the particle
distributions of substructure in the Aquarius simulations (assigned to meshes
according to the CIC algorithm) to the isothermal ellipsoid that fitted to the
main galaxy halo, and compare its lensing properties with those of the smooth
underlying isothermal ellipsoid. In this approach, the analytical solutions of
the fitted isothermal ellipsoidal potential and its derivatives are tabulated
on the image grid, including the cusp-caustic relation. By doing so, no
Poisson noise of the underlying main halo is introduced, and so any confusion
to the results from substructures is avoided.

We fit an isothermal ellipsoidal model to each of the three independent
projections of each galaxy halo. The six fitting parameters are: (1) the
effective critical radius $b_{\rm I}$, (2) the axis-ratio of the surface
density ellipse $q_3$, (3) the core radius $S_0$, (4)-(5) the $X$- and
$Y$-offsets of the projected centre $X_{\rm c}$, $Y_{\rm c}$, and (6) the
rotation angle of the major axis of the surface density ellipse $RA$. The
uncertainties of the fitted parameters are $\Delta b_{\rm I} =
0.003\arcsec$, $\Delta q_3 = 0.01$, $\Delta S_0 = 0.001$, $\Delta X_{\rm c} =
\Delta Y_{\rm c} = 0.002\arcsec$, $\Delta RA = 0.004\pi$. The relative errors
in the fitted critical curves and caustics are $\lesssim$ 10 \% for
different projections of all the simulated galaxy haloes.

In Table \ref{tab:Aquarius_HalosCusp}, we list the isothermal
ellipsoid parameters of the main haloes ($b_{\rm I}$, $q_3$ and
$S_0$). The critical radius $b_{\rm I}$ is of the order of
$0.3\arcsec$ to $0.9\arcsec$. The separation between images ($\sim 2
b_{\rm I}$) is in the range of the observed gravitational lenses
(which peaks around $1\arcsec$, see e.g. \citealt{Browne2003}). The
axial ratios also match the observed lenses quite well. There is one
exception, however. The core radius $S_0$ is quite large, of the
order of ($0.05\arcsec - 0.1\arcsec$, corresponding to a few
hundreds of pc). Such a core size is larger than those inferred 
from gravitational lenses which are in general consistent with zero core
radius (e.g. \citealt{WN1993, RusinMa2001, Oguri2001, LiOstriker2003}). 
This is a direct result of the implementation of the Hernquist profiles, 
which follow a logarithmic density slope of $-1$ in the central regions 
(see Fig.~\ref{fig:Mainhalo1}). However, this artifact should have little
effects on the images we are interested in, which are close to the
outer critical curve.

\begin{figure*}
\centering
\includegraphics[scale = 1.0]{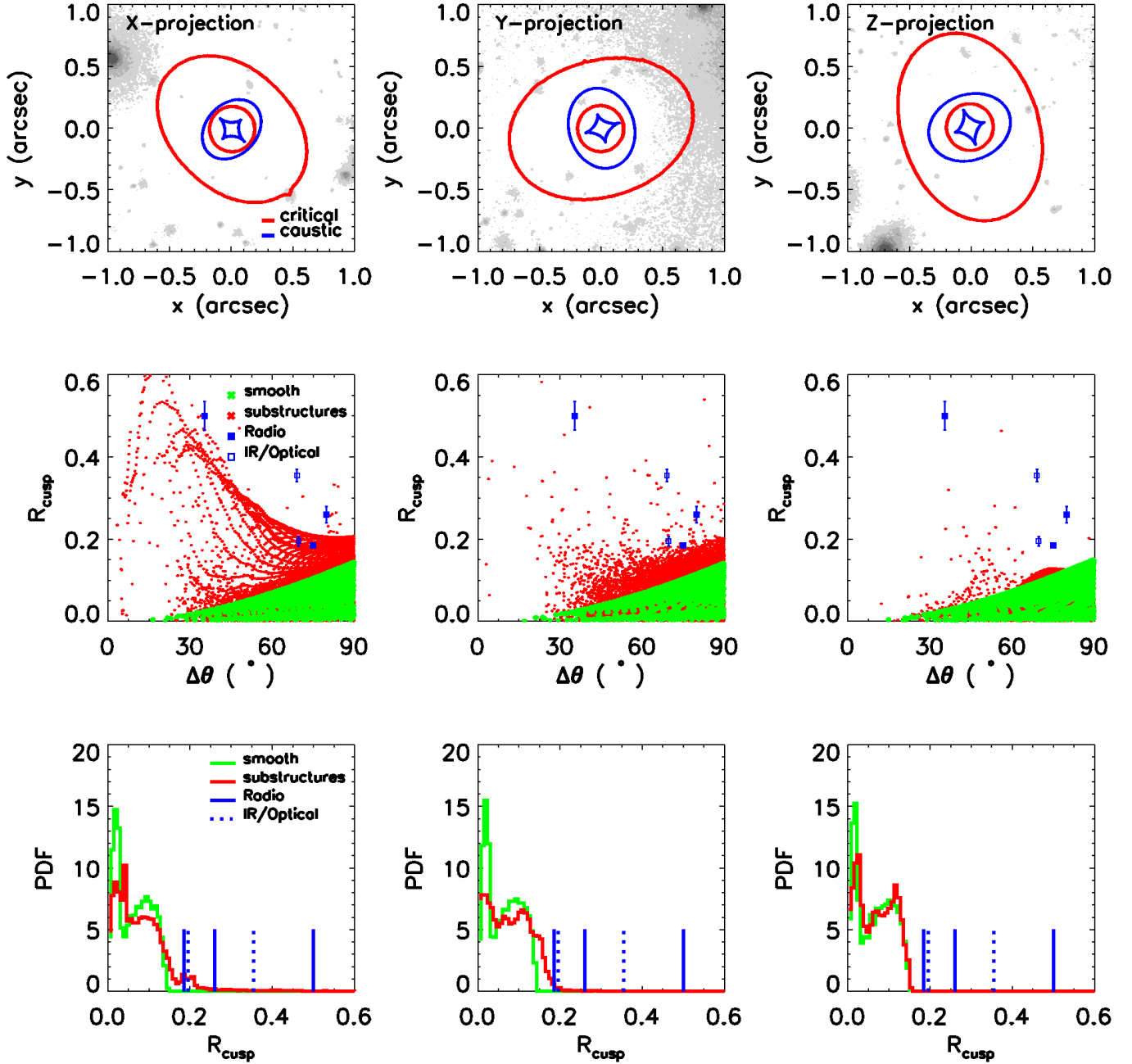}
\caption{Lensing properties for the halo {\it Aq-A-2}, in three independent
  projections. The top panels show the critical curves (red) and caustics
  (blue) superimposed on top of the subhalo population. The middle panels show
  $\Rcusp$ vs. the image opening angle $\Delta\theta$. Large $\Rcusp$ values
  (red) are due to substructure. The triangle pattern (green) gives
  predictions for the smooth counterparts. The bottom panels show the
  corresponding probability distribution functions (PDFs) of $\Rcusp$ for cusp
  sources with $\Delta\theta \leq 90 ^{\circ}$. The violation of the
  cusp-caustic relation can be seen from the excess of $\Rcusp$ at large
  values (red) over the smooth counterpart curve (green). The $\Rcusp$ values
  for the three radio and two optical/IR cusp lenses are indicated by vertical
  solid and dashed bars (see Table
  \ref{tab:ObservedRcusp}).} \label{fig:HaloRcusp_C02}
\end{figure*}

\begin{figure*}
\centering
\includegraphics[scale = 1.0]{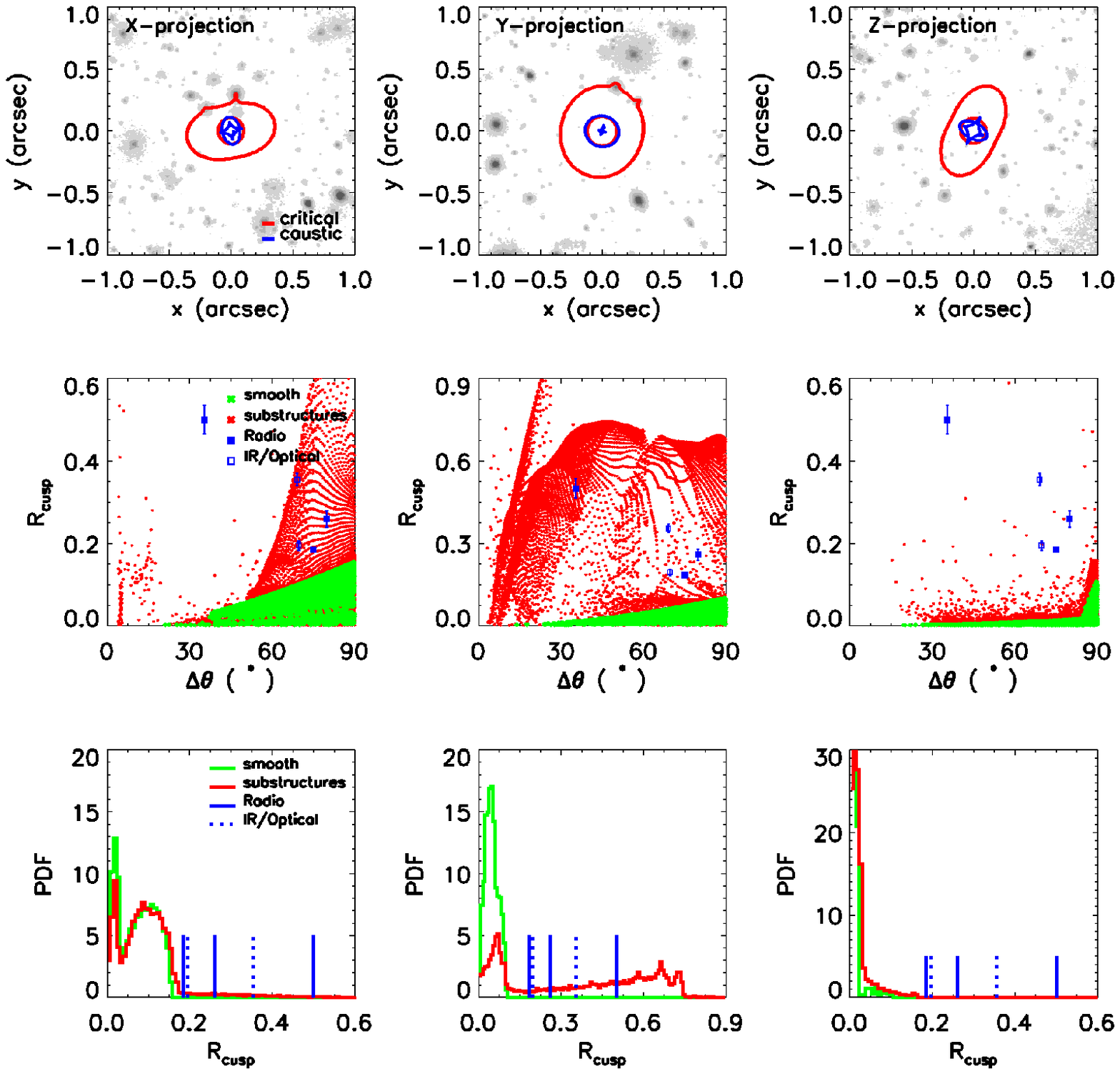}
\caption{For the halo {\it Aq-B-2}, the symbols are the same as in
  Fig.~\ref{fig:HaloRcusp_C02}. The truncated triangle pattern in the
  $Z$-projection is due to naked cusps of the central caustic. The strong
  violation of the cusp-caustic relation seen in the $Y$-projection is caused
  by subhaloes with $\msub \leq 10^8 h^{-1}M_{\odot}$ with a violation rate
  $P(\Rcusp \ge 0.187)=64\%$. }
\label{fig:HaloRcusp_C06}
\end{figure*}

\begin{figure*}
\centering
\includegraphics[scale = 1.0]{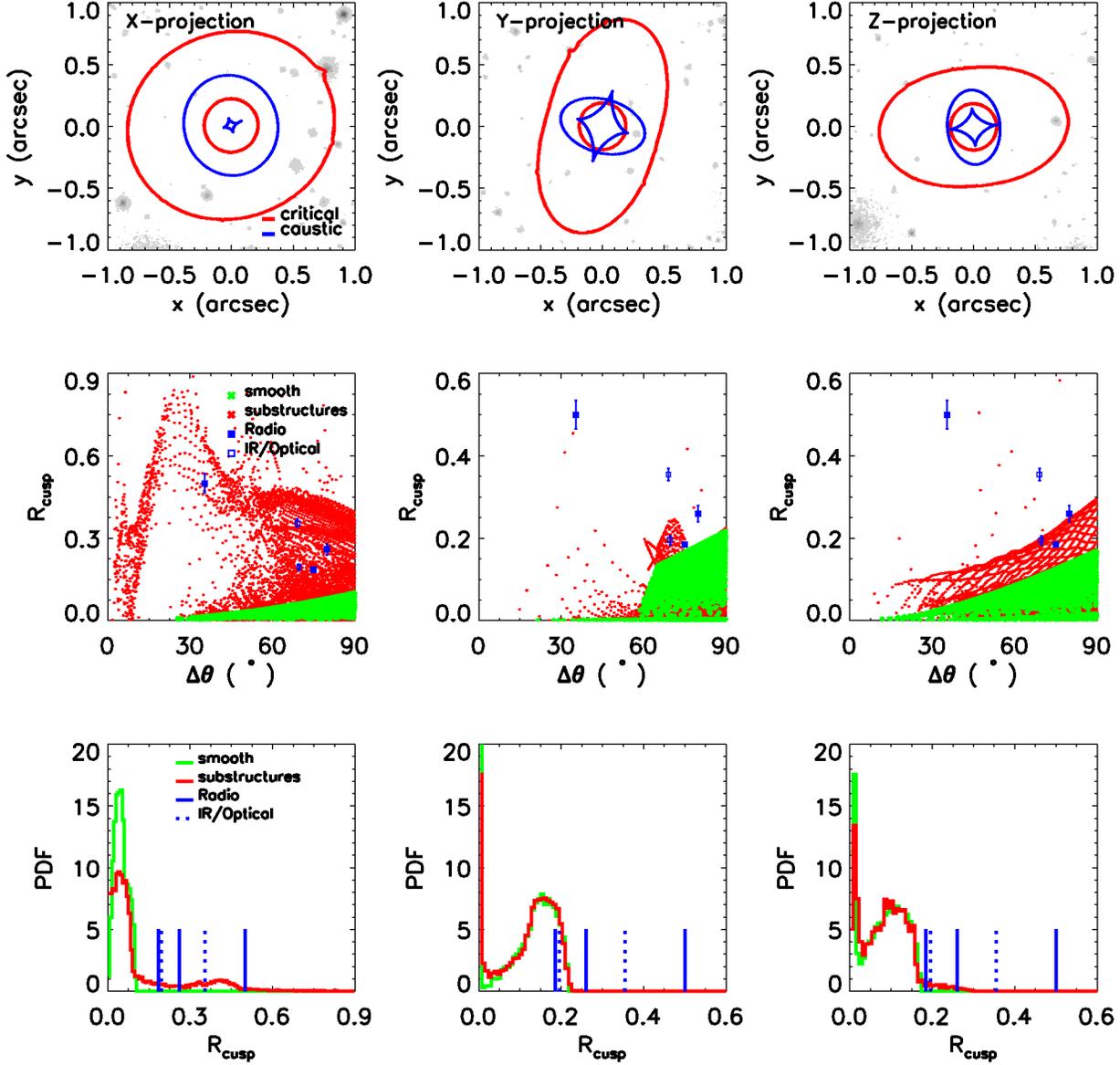}
\caption{For the halo {\it Aq-C-2},  the symbols are the same as in
Fig.~\ref{fig:HaloRcusp_C02}.  The truncated triangle pattern in the
$Y$-projection is due to naked cusps of the central caustic. The
halo in this projection has large ellipticity, which results in
large $\Rcusp$ values. The strong violation in the $X$-projection is
caused by subhaloes with $\msub \leq 10^8 h^{-1}M_{\odot}$ with a
violation rate $P(\Rcusp \ge 0.187)=19\%$. }
\label{fig:HaloRcusp_C09}
\end{figure*}

\begin{figure*}
\centering
\includegraphics[scale = 1.0]{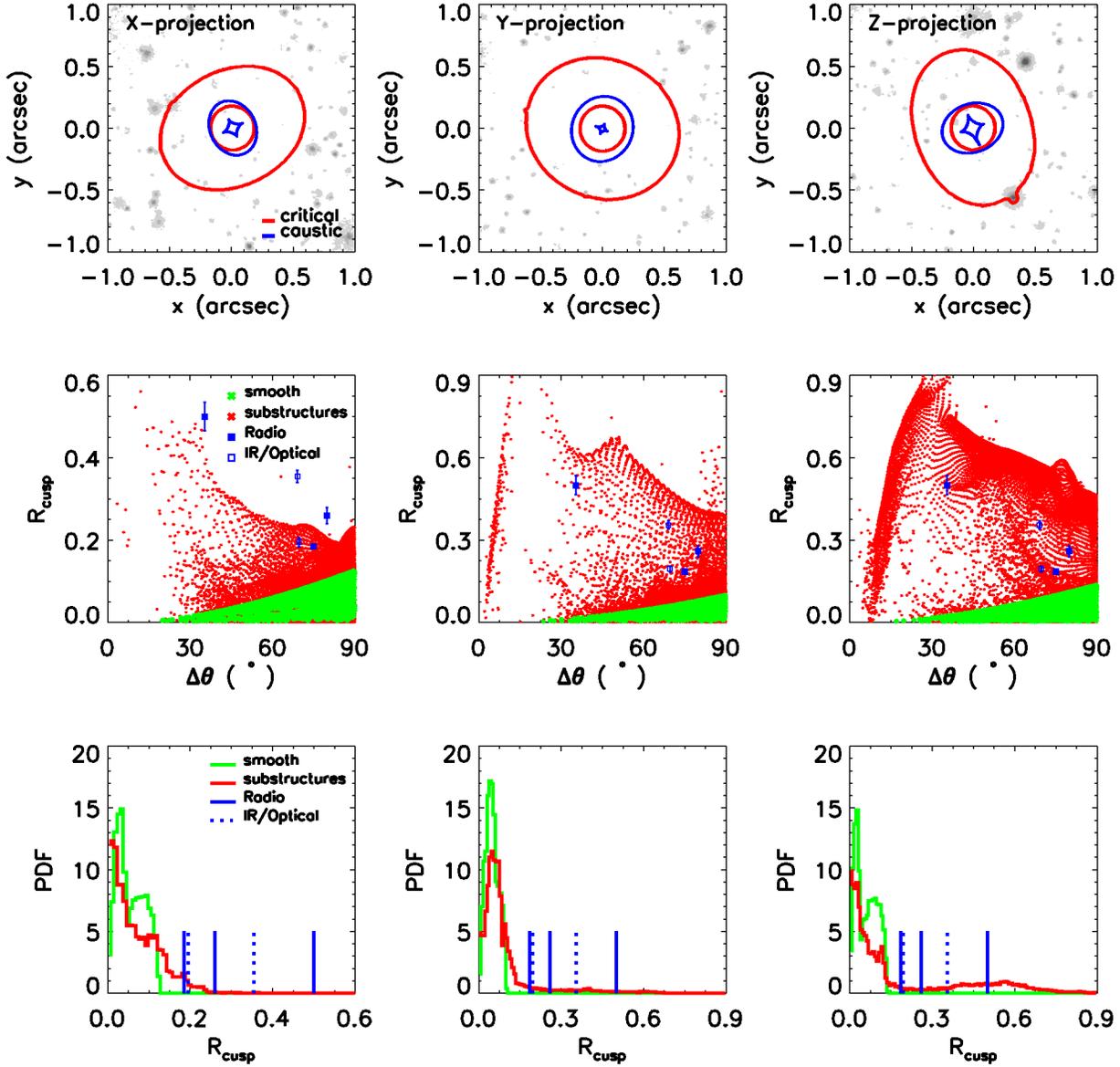}
\caption{For the halo {\it Aq-D-2}, the symbols are the same as in
Fig.~\ref{fig:HaloRcusp_C02}. The strong violations in the $Y$- and
$Z$-projection are caused by subhaloes with $\msub \leq 10^7
h^{-1}M_{\odot}$ and $\msub \leq 10^8 h^{-1}M_{\odot}$, respectively,
with violation rates $P(\Rcusp \ge 0.187)=9.7\%$ ($Y$-projection)
and $31\%$ ($Z$-projection).  } \label{fig:HaloRcusp_91439}
\end{figure*}

\begin{figure*}
\centering
\includegraphics[scale = 1.0]{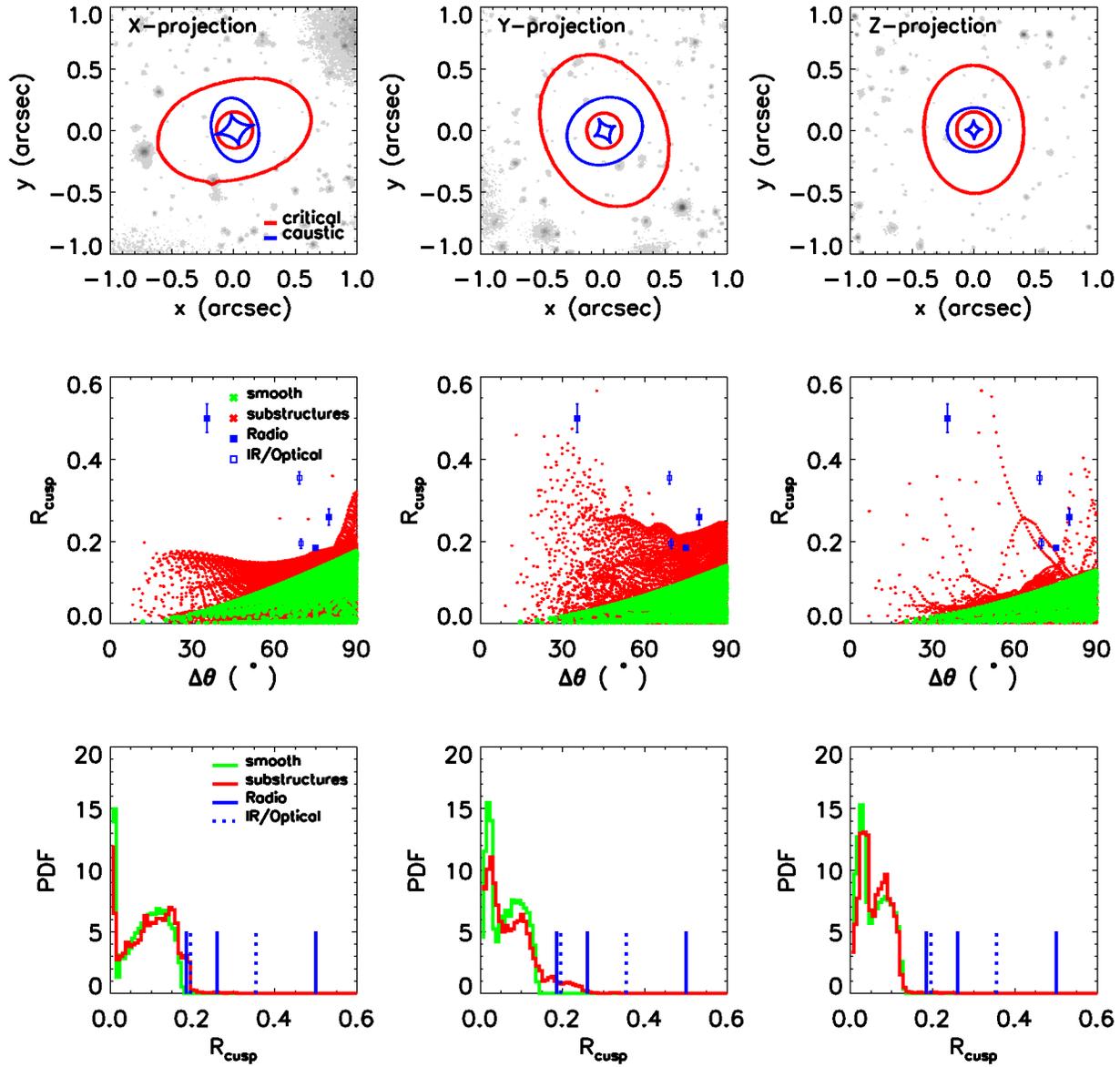}
\caption{For the halo {\it Aq-E-2}, the symbols are the same as in
  Fig.~\ref{fig:HaloRcusp_C02}.}
\label{fig:HaloRcusp_9153}
\end{figure*}

\begin{figure*}
\centering
\includegraphics[scale = 1.0]{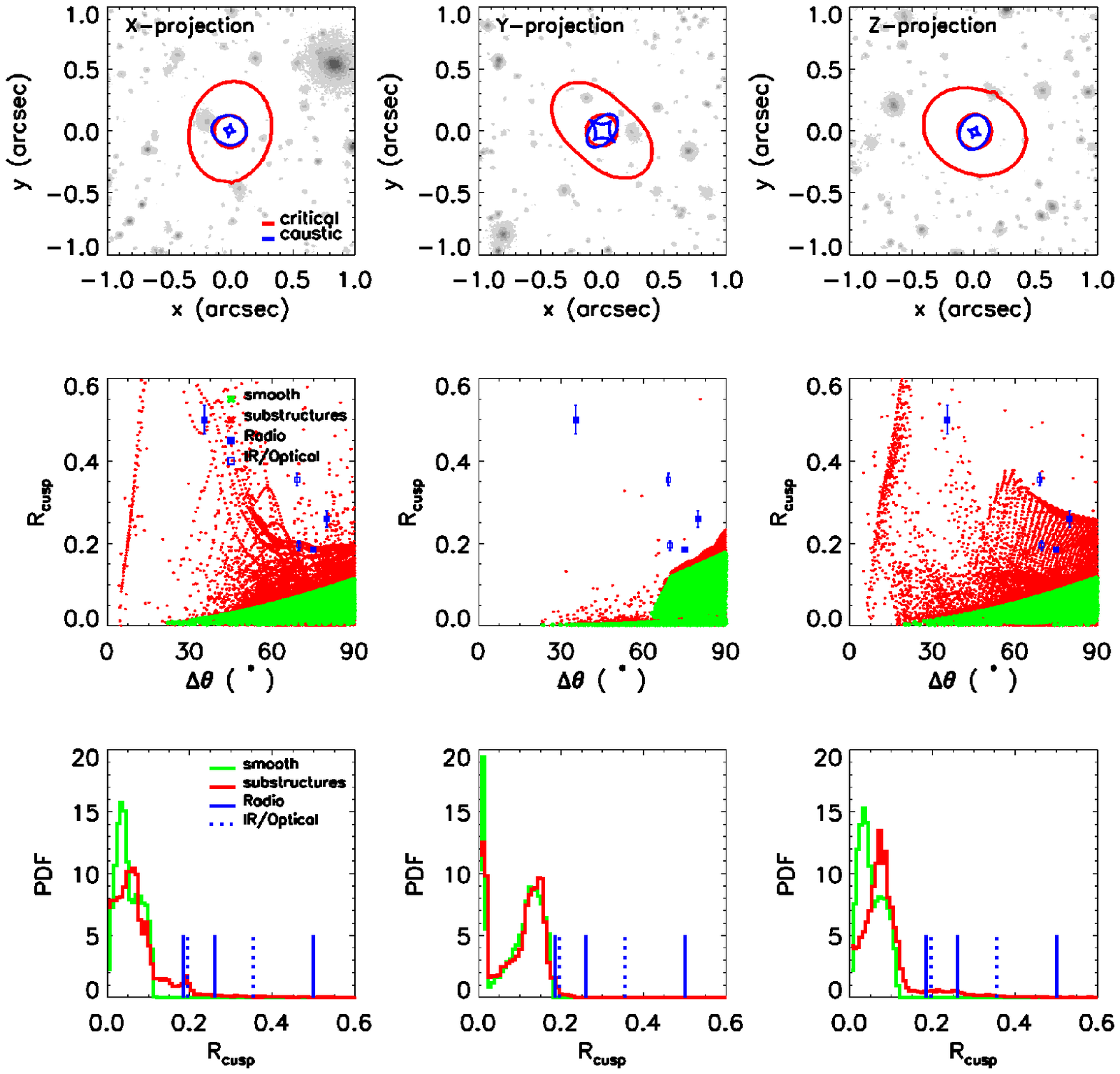}
\caption{For the halo {\it Aq-F-2}, the symbols are the same as in
  Fig.~\ref{fig:HaloRcusp_C02}. The truncated triangle pattern in the
  $Y$-projection is due to naked cusps of the central caustic. The strong
  violation in the $Z$-projection is mainly caused by subhaloes with
  $\msub \leq 10^7 h^{-1}M_{\odot}$ with
  a violation rate $P(\Rcusp \ge 0.187)=6.7\%$. }
\label{fig:HaloRcusp_9470}
\end{figure*}

\begin{figure*}
\centering
\includegraphics[scale = 1.0]{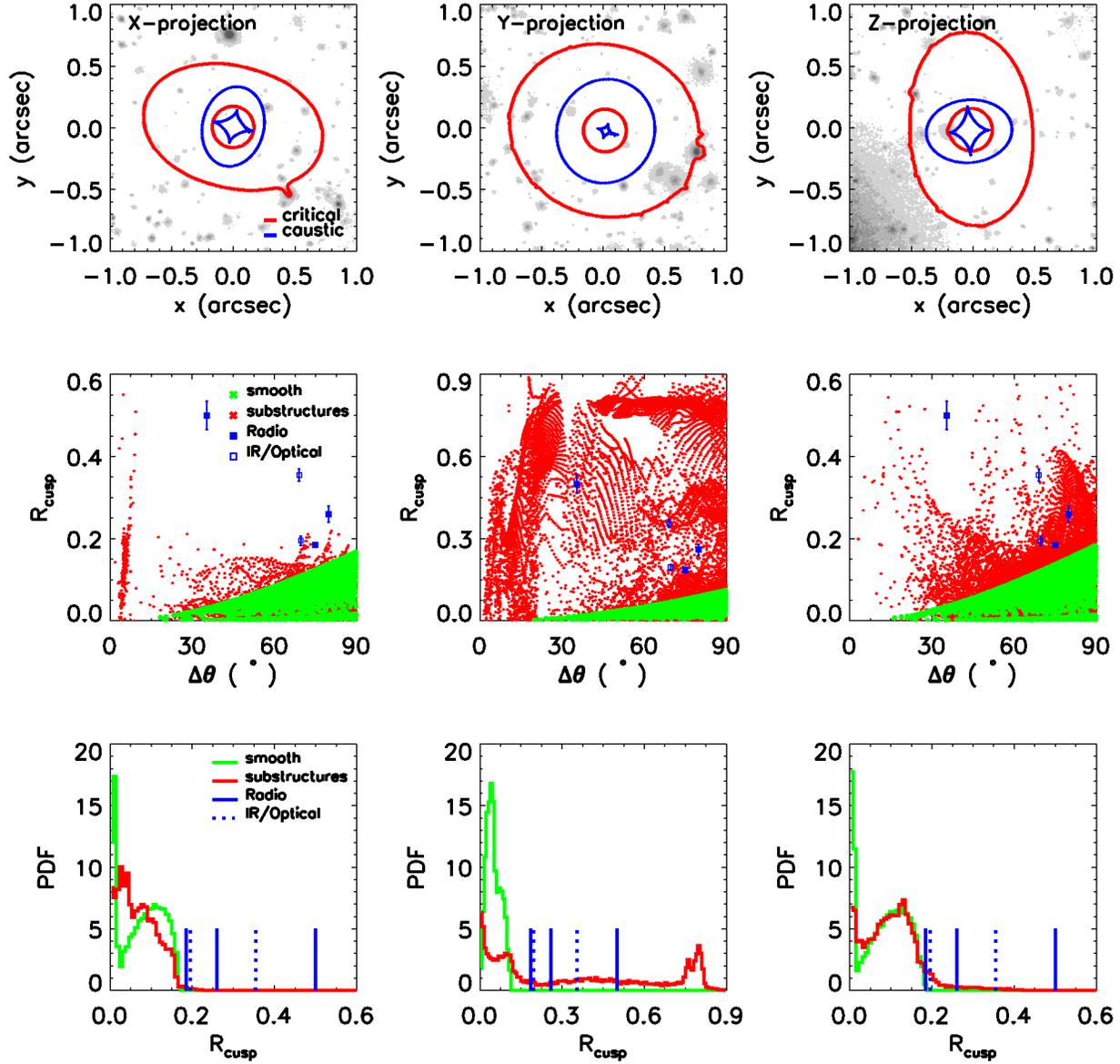}
\caption{For the halo {\it Aq-A-2} at $z$=0.6, the symbols are the
same as in Fig.~\ref{fig:HaloRcusp_C02}. The strong violation in the
$Y$-projection is caused by subhaloes with $\msub \leq 10^8
h^{-1}M_{\odot}$; a cusp violation rate is $P(\Rcusp \ge
0.187)=56\%$. } \label{fig:HaloRcusp_C02Z0p6}
\end{figure*}

To examine the violation of the cusp-caustic relation, we generate
about 10000 cusp sources in each case, and calculate the resulting
$\Rcusp$ distributions. All these cusp sources are inside the
central caustic and close to the cusps, where the corresponding
triple images have opening angles $\Delta\theta \leq 90^{\circ}$.
The results of all 7 studied Aquarius haloes (in 21 projections) are
given in Fig.~\ref{fig:HaloRcusp_C02} to
Fig.~\ref{fig:HaloRcusp_C02Z0p6}. As mentioned above, the
probability density distribution of $\Rcusp$ often shows two peaks
for the smooth haloes which are produced by the major and minor
cusps, respectively. For the ``naked'' cusp cases (where the central
diamond caustic protrudes the outer elliptical caustic), the
distributions of $\Rcusp$ vs. the opening angle $\Delta\theta$ are
somewhat truncated below certain opening angles (see
Fig.~\ref{fig:HaloRcusp_C09} for the halo {\it Aq-C-2}'s
$Y$-projection for an example). Empirically, it is rare for
massive lensing galaxies to produce naked cusps. There is only one
candidate APM08279 (\citealt{Lewis2002NakedCuspAPM08279}), and that
is likely due to lensing by an edge-on spiral rather than an
elliptical. The four naked cusp cases from our simulations are caused 
by the large cores in the central density profiles of the lensing
galaxies. We exclude these four naked cusp cases in the final statistic 
calculations (their inclusion does not significantly alter our results).
Strong violations of the cusp-caustic relation due to substructures
are seen in some cases, e.g. for the $Y$-projection of the {\it
  Aq-B-2} halo (see Fig.~\ref{fig:HaloRcusp_C06}). However, most of
these cases have small cusp-lensing cross-sections (listed in Table
\ref{tab:Aquarius_HalosCusp}, Column 8), defined as the areas
covered by cusp sources whose images satisfy $\Delta\theta \leq 90
^{\circ}$. The mean probability of cusp violations calculated below
are weighted by the cross-sections (see eq. \ref{eq:psigma}). As can
be seen from Fig.~\ref{fig:HaloRcusp_C02} to
Fig.~\ref{fig:HaloRcusp_9470}, the scatter in the cusp violation is
large between different projections of different haloes. Also notice
that the halo {\it Aq-A-2} at $z=0.6$
(Fig.~\ref{fig:HaloRcusp_C02Z0p6}) does not show a significant
difference from the redshift-zero haloes in the violation of the
cusp-caustic relation.

\begin{figure*}
\centering
\includegraphics[scale = 0.6]{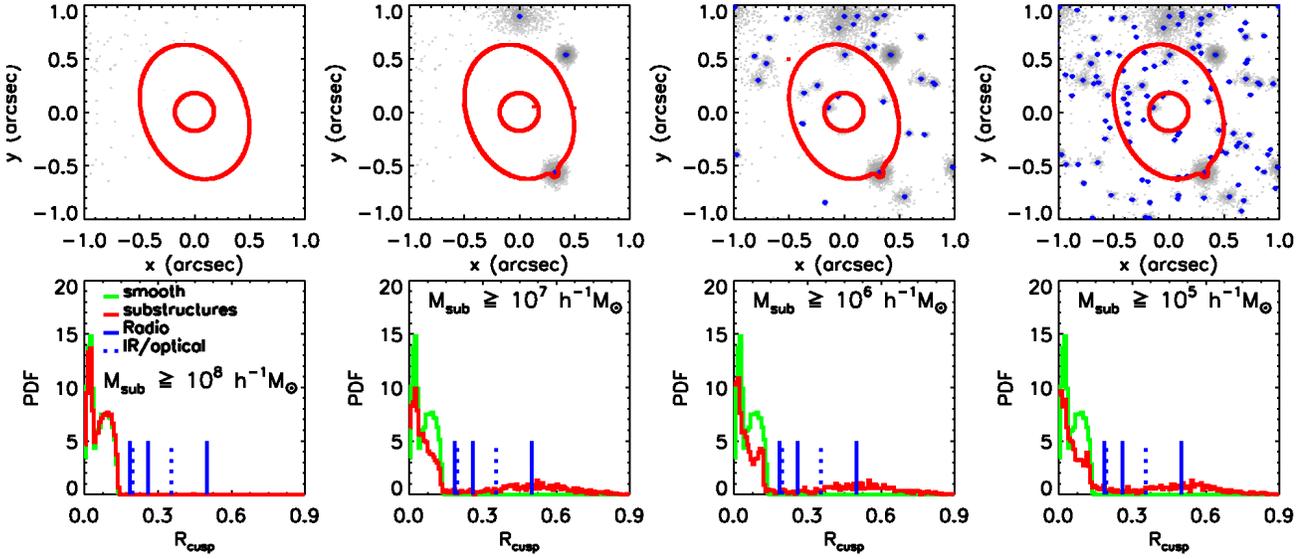}
\caption {Effects of substructure lensing as a function of the lower cutoff
  subhalo mass for the halo {\it Aq-D-2} along the $Z$-projection. The upper
  panels show the projected substructures with masses above a threshold. The
  projected centres of the subhaloes and the corresponding critical curves are
  plotted at the top.  From the left to the right, the lower cutoff subhalo
  mass changes from $10^8 h^{-1}M_{\odot}$, $10^7 h^{-1}M_{\odot}$, $10^6
  h^{-1}M_{\odot}$ to $10^5 h^{-1}M_{\odot}$. The bottom panels show the
  corresponding probability distribution functions of $\Rcusp$. Most
  substructures that survive and are projected within the central few kpc are
  low-mass subhaloes ($\leq 10^8$ $h^{-1}M_{\odot}$), which dominate the
  violation of the cusp-caustic relation.}
\label{fig:MassConvergeExample}
\end{figure*}

To see which massive substructures cause the cusp-caustic violation, we
calculate the $\Rcusp$ distribution due to subhaloes more massive than $10^5
h^{-1}M_{\odot}$, $10^6 h^{-1}M_{\odot}$, $10^7 h^{-1}M_{\odot}$, and $10^8
h^{-1}M_{\odot}$, respectively.
Fig.~\ref{fig:MassConvergeExample} shows one typical
example, for the halo {\it Aq-D-2} along the $Z$-projection. We find that in
most cases substructures with
masses $\msub \leq 10^7$ to $10^8 h^{-1}M_{\odot}$
dominate the contribution to the violations of the cusp-caustic relation (see
the Col (9): $M_{\rm sub, cr}$ in Table \ref{tab:Aquarius_HalosCusp}).
Notice that previous studies on cusp violations typically resolve haloes
larger than $\sim 10^8 h^{-1} M_\odot$, and thus would not have been able to
evaluate the effects of substructure accurately.  However, the addition
of subhaloes with $\msub \la 10^6 h^{-1} M_\odot$ does not appear to
increase the violation frequency significantly (compare the three right
panels). We return to the convergence issue as a function of subhalo mass
in \S\ref{sec:discussion}.

\begin{figure*}
\centering\includegraphics[scale=0.7,angle=90]{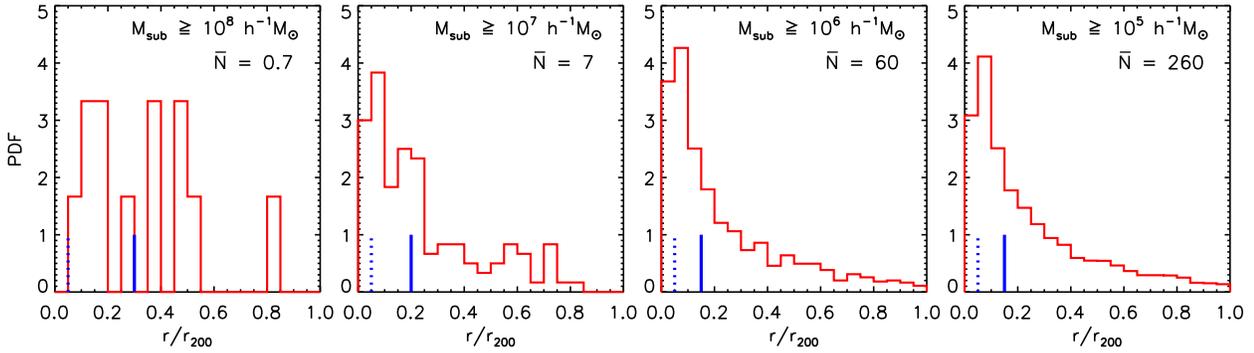}
\caption{Distribution of halocentric distances of the subhaloes projected
  within 0.05 $r_{200}$ ($\sim$ 2.5 times the Einstein radius, indicated by
  the dotted line in each panel). The solid lines give the median spherical
  halocentric distances of the subhaloes; all are around 0.2 $r_{200}$. The
  average number of subhaloes $\bar{N}$ is indicated inside each panel.  }
\label{fig:Subhalo_distance}
\end{figure*}

Notice that most subhaloes that are projected close to the critical
curves are due to chance alignment. Fig.~\ref{fig:Subhalo_distance} shows
the spherical halocentric distance distribution for the subhaloes that are
within a projected distance of 0.05 $r_{200}$ ($\sim 2.5$ Einstein
radii). The fractions of subhaloes that are physically located within a
spherical radius of 0.05 $r_{200}$ are 15\%, 18\%, 15\% and 0\% for
subhaloes more massive than $10^5 h^{-1}M_{\odot}$, $10^6 h^{-1}M_{\odot}$,
$10^7 h^{-1}M_{\odot}$ and $10^8 h^{-1}M_{\odot}$, respectively. The
large median halocentric distances, $\sim 0.2$ $r_{200}$ in all cases,
also show that projection effects are substantial.

\subsection{Comparison with observations} \label{sec:comparison}

\citet{KGP2003apj} summarised 19 published quadruply imaged systems. Seven of
them are detected at radio wavelengths\footnote{B0128+437
  (\citealt{Phillips2000B0128}), B0712+472 (\citealt{Jackson98B0712};
  \citealt{Jackson2000B0712}), B1422+231 (\citealt{Impey1996B1422};
  \citealt{Patnaik2001B1422}), B1555+375 (\citealt{Marlow1999B1555}),
  B1608+656 (\citealt{KoopmansFassnacht1999B1608}), B1933+503
  (\citealt{Cohn2001B1933}) and B2045+265
  (\citealt{Fassnacht1999B2045}).}.
Radio lenses are free from dust extinction. 
Due to their large emission regions, they are less likely to be affected by
microlensing. In contrast, microlensing is likely to affect optical/IR flux
ratios and so we treat them differently below. 

\citet{DK2002} studied seven four-image radio-lensing systems:
MG0414+0534 (\citealt{Hewitt1992MG0414}), B0712+472
(\citealt{Jackson98B0712}), PG1115+080
(\citealt{Weymann1980PG1115}), B1422+231
(\citealt{Patnaik2001B1422}), B1608+656
(\citealt{Fassnacht1996B1608}), B1933+503 (\citealt{Sykes1998B1933})
and B2045+265 (\citealt{Fassnacht1999B2045}) and found that six show
anomalous flux ratios, which might be due to the effects of
substructure lensing. Among all the detected radio lenses, three
(B0712+472, B1422+231 and B2045+265) show a typical cusp-caustic
geometry (with $\Delta\theta \leq 90 ^{\circ}$) and violations of
the cusp-caustic relation. Another two lensing systems observed in
the optical/IR band are also cusp-caustic lenses with $\Delta\theta
\leq 90 ^{\circ}$: RXJ1131-1231 (\citealt{Sluse2003aaJ1131}) and
RXJ0911+0551 (\citealt{Bade1997aa317, Burud1998apjl}). Both
have unexpected large values of $\Rcusp$, which were shown to have
been affected by microlensing (\citealt{Morgan2006RXJ1131,
Anguita2008Microlensing}). Table \ref{tab:ObservedRcusp} lists the
$\Rcusp$ and $\Delta\theta$ values for the five observed
cusp-caustic lenses. Three out of the five cusp lenses are
detected at radio wavelengths, thus their large $\Rcusp$ values are
unlikely due to microlensing. We treat these three radio lenses as
cusp-caustic violations due to substructure lensing. 
Below we will calculate the probability for
the simulations to reproduce such an observed violation rate.

For each galaxy and each projection, we calculate the violation probability
that the predicted $\Rcusp$ is larger than the observed $\Rcusp$ value 0.187
for B1422+231, which shows the smallest violation (smallest $\Rcusp$ value)
among the five cusp lenses with $\Delta\theta \leq 90 ^{\circ}$. The
cross-section weighted violation probability is given by
\begin{equation}
p_\sigma = \sum_{i} f_{\sigma, i} \,
p_{i}(\Rcusp \geq 0.187 |\Delta \theta \leq 90^\circ),
f_{\sigma, i} = \frac{\sigma_i}{\sum_{i} \sigma_{i}},
\label{eq:psigma}
\end{equation}
where the summation $i=1, \cdot\cdot\cdot, (21-4)$ is for
the seven haloes along the three independent projections of each,
excluding the four naked cusp cases, and $\sigma_i$ is the
cross-section in the source plane for producing three close images
with opening angle $\Delta\theta \leq 90 ^{\circ}$. Using the above
formula, we find the mean probability $p_\sigma \approx
6.4\%$ for $\Rcusp \geq 0.187$. Notice that this probability
estimate is only approximate, since we have not considered the
magnification bias (e.g. \citealt{TOG}).

\begin{table}
\centering
\caption{The image opening angle and $\Rcusp$ for the observed
  cusp-caustic lenses, taken from \citet{AB06mn}.}
\label{tab:ObservedRcusp}
\begin{minipage} {\textwidth}
\begin{tabular}[b]{l|l|c|l}\hline
\tableline Lens & $\Delta\theta$ & $\Rcusp$ & Band \\\hline
\tableline B0712+472 & 79.8$^{\circ}$ & 0.26 $\pm$ 0.02 & radio \\
\tableline B2045+265 & 35.3$^{\circ}$ & 0.501 $\pm$ 0.035 & radio \\
\tableline B1422+231 & 74.9$^{\circ}$ & 0.187 $\pm$ 0.006 & radio \\
\tableline RXJ1131 - 1231 & 69.0$^{\circ}$ & 0.355 $\pm$ 0.015 & optical/IR \\
\tableline RXJ0911 + 0551 & 69.6$^{\circ}$ & 0.192 $\pm$ 0.011 & optical/IR \\
\hline
\end{tabular}
\end{minipage}
\end{table}

To have three (radio) lensing cases with $\Rcusp \geq 0.187$ 
(due to substructure lensing rather than microlensing)
out of the five cusp lenses ($\Delta\theta \leq 90^\circ$) observed
so far, the probability is $C_5^3p_\sigma^3(1-p_\sigma)^2 \approx
2.3\times 10^{-3}$. The low probability suggests that the subhalo
populations in the inner regions of the Aquarius haloes with
Hernquist galaxies are insufficient to explain the observed
frequency of flux anomalies in the cusp lenses.

\section{DISCUSSION AND CONCLUSIONS}
\label{sec:discussion}

In this paper, we have used the ultra-high resolution Aquarius simulations to
study the effects of substructure lensing. We incorporate the effects of
baryons in the main halo by adding a stellar component (modelled as a
Hernquist profile), and then take into account its effects on the dark matter
halo through adiabatic contraction. The density profiles and lensing
properties except the flux ratios are broadly consistent with the observed
gravitational lenses. Using Monte Carlo simulations, we find large numerical
noise for an isothermal halo populated with $10^8$ particles, which shows
considerable scatter in the $\Rcusp$ distribution for cusp lenses. In the end,
we therefore study the substructure lensing by modelling the smooth underlying
galaxy halo as an isothermal ellipsoid and superimposing the subhalo
population from the Aquarius simulations.  In this way, we focus on the
lensing effects of subhaloes and avoid any confusion from numerical noise in
the $N$-body realisation of the simulated main haloes.

Our study finds that even with the much better resolved subhalo
population of the Aquarius simulations, the observed cusp lenses
still violate the cusp-caustic relation more frequently than
predicted by $N$-body simulations.

The Aquarius haloes are Milky Way type haloes in terms of their
masses, while many lenses are ellipticals, which are more massive.
Among the five cusp lenses we compare our results with, three of them
(B2045+265, RXJ1131-1231, RXJ0911) are more massive than our simulated
haloes and have Einstein radii twice as large as those of our haloes.
The other two lenses (B0712+472, B1422+231) have Einstein radii and
velocities (circular velocity or velocity dispersion) roughly comparable
to the relatively massive haloes in the Aquarius simulations.
As shown in Fig.~\ref{fig:Subhalo_properties} (the right panel) the
projected subhalo mass fraction increases with the projected radius $R$. 
If we have under-estimated the Einstein radii $b_I$ (e.g. because 
of uncertainties in the addition of the central galaxies), 
we could have potentially under-estimated the violation
rates due to the lack of enough substructures at smaller radii.
We artificially increase the Einstein radii of the simulated haloes
by a factor of two to study the violation probabilities
due to a higher fraction of substructures at larger radii.
The mean subhalo mass fraction within a $0.1\arcsec$-annulus around
the new Einstein radius would increase from $f_{\rm sub, annu}$ $\approx$
0.19\% to 0.24\%, and the mean violation probability would increase
from $p_\sigma \approx$ 6.4\% to 14.0\%. The probability of reproducing
the observed violation rate would increase from 0.2\% to 2\%.

Another concern is that due to the finite particle mass in
$N$-body simulations, the central cusps of the subhaloes may not be
resolved, which may potentially result in an under-estimation of the
ability of the subhaloes to induce perturbations to the lensing
potential. We consider an extreme case assuming all subhaloes
are point-like sources with their masses and locations from the
simulations. In this scenario, $f_{\rm sub, annu}$ roughly
remains at 0.18\%, however, $p_\sigma$ increases to 15.1\%. 
The probability to reproduce the observed violation rate increases to 2.5\%.

These low probabilities suggest that the subhalo populations
in the central regions of the Aquarius haloes are not sufficient to
explain the observed frequency of violations of the cusp-caustic
relations. It is important to ask whether our results will change
significantly if even lower-mass subhaloes are resolved. We argue
that this is unlikely to be the case.  The total subhalo lensing
cross-section is an integral of the cross-section of subhaloes of
each mass weighted by their abundance. As shown in
\S\ref{sec:aquarius}, most of the perturbing subhaloes have
relatively low mass ($\msub \leq 10^7$ to $10^8 h^{-1}M_{\odot}$).
Their abundance scales as ${\rm d}N(\msub)/{\rm d}\msub \propto
\msub^{-1.9}$.  For a galaxy (subhalo) approximated by a SIS, the
lensing cross-section roughly scales as $\sigma^4$ (e.g.
\citealt{TOG}) where $\sigma$ is the one-dimensional velocity
dispersion. For Aquarius subhaloes, $\msub \propto V_{\rm max}^{3}$
(\citealt{volker08Aq}), where $V_{\rm max}$ is the maximum circular
velocity. If $\sigma \propto V_{\rm
  max}$, then the integrated lensing cross-section will be $\propto
\msub^{0.43}$.  On the other hand, for a point lens or an elliptical galaxy,
the lensing cross-section is proportional to the lens mass, and the
integrated lensing cross-section would be $\propto \msub^{0.1}$.  In all
these cases, the subhalo lensing cross-sections are biased towards
relatively massive subhaloes in the projected central region, and the
incorporation of even lower mass subhaloes should not change our results
significantly.

We mention in passing that a warm dark matter scenario would suppress the
formation of small subhaloes, making it even more difficult to explain the
observed cusp violations (see e.g.  \citealt{Miranda2007}). Below, we compare
our study with previous work, before discussing its limitations and outlining
possible future work.

\subsection{Comparison with previous studies}

There have been a number of studies of substructure lensing using numerically
simulated haloes, including those from hydrodynamical simulations. Below we
compare a few of these studies with our own.

\citet{DK2002} concluded that at the 90\% confidence level, a substructure
fraction of 0.6\% to 7\% can explain the observed anomalous flux ratio. For
the Aquarius subhalo population, Table \ref{tab:Aquarius_HalosCusp} Col (5):
$f_{\rm sub, annu}$ shows such fraction averaged over a thin annulus around
the outer tangential curve, which is always below 1\%, sometimes much
smaller (not to be confused with $f_{\rm sub}$ in
Table \ref{tab:Aquarius_simulations} and Fig.~\ref{fig:Subhalo_properties},
which refers to the subhalo mass fraction within $r_{200}$).
This is the primary reason why our predicted cusp violations are
smaller than the observed violation frequencies.

\citet{BS2004aa} used hydrodynamical simulations of
\citet{Steinmetz2002newastro} and concluded that the predicted cusp violations
due to substructure are comparable to those observed. Their simulated halo
has $\sim 10^5$ particles, resolving subhaloes down to $5\times 10^8
M_\odot$. As the authors pointed out, the numerical noise may be as high as
5\%. The observed high-order singularities in their simulations are much
higher than ours (comparable to the caustic structure shown in
Fig.~\ref{fig:MCIE_Rcusp1} for $10^6$ particles). It is possible that
their high numerical noise may have produced too many artificial
violations, although we note that they used Voroni density estimation
to reduce the discreteness noise.

Our conclusion that the dark matter subhalo population may be 
insufficient to explain the observed cusp violations is consistent 
with \citet{MaoJing04apj}, \citet{AB06mn}, \citet{Maccio2006b} 
and \citet{Maccio2006}. The number of particles used in those studies 
is roughly two orders of magnitude smaller than here. In particular, 
the study by \citet{MaoJing04apj} found large scatter among different 
haloes, a conclusion confirmed by our results. 


\subsection{Limitations of the present study and future work}

The most severe limitation of our study is that the high-resolution
simulations used here include only dark matter. Without baryons,
these haloes are sub-critical (see \S\ref{sec:darkLight}) and
incapable of producing multiple images. We are therefore forced to
incorporate a model for the baryonic galaxy at the centre of each
halo. The galaxy changes not only the overall dark matter profiles
(taken into account by adiabatic contraction), but also the
dynamical evolution of subhaloes, an effect which is not considered
here. On the one hand, the increased baryonic density at the centre
of the halo will make the subhaloes feel stronger tidal forces,
particularly those that come close to the centre. On the other hand,
the baryons within subhaloes will make them more resilient to tidal
disruption.  It is not clear which effect will dominate. We comment,
however, that the subhaloes that come very close to the centre may
have already been tidally stripped or disrupted, and thus most of
the surviving subhaloes that can be identified by {\tt SUBFIND} may
have quite large peri-centre passages. As a result, the effects of
baryons in the host halo may not change the results very
significantly. However, we caution that, SUBFIND, like most
substructure finders, has difficulties in identifying subhaloes in
the densest regions of the halo and assigning them correct masses.
Empirically the Milky Way does not seem to host many luminous 
satellites close to the centre.
Hydrodynamical simulations can in principle address this issue
directly (subject to the uncertainties in the treatment of gas
processes). \citet{Maccio2006b} found a factor of two increase in
the number of surviving satellite galaxies (with masses 
above $10^7 M_{\odot}$) in the centres of galaxies when including baryons 
in the CDM simulations, but concluded that even this was not sufficient to 
explain the flux anomaly problem.

Observationally, it is interesting that more than one half of the
CLASS lenses appear to show luminous companion galaxies in
projection (\citealt{BMK2008, Jackson2009}), and their inclusion in
the models appears to alleviate the anomalous flux ratio problem
(see below). This may be just a statistical fluke due to the small
sample size (22 lenses in total) or some of these may be due to
chance alignment along the line-of-sight
(\citealt{Chen2003,Wambsganss2005,Metcalf2005a}a,b;
\citealt{Miranda2007}). Nevertheless, for the three radio lenses
that show apparent cusp violations (see Table
\ref{tab:ObservedRcusp}), the most serious case is B2045+265 with
$\Rcusp\approx 0.5$. Recently, \citet{McKean2007} found a galaxy,
G2, which is about 0.66 $\arcsec$ away from the main lensing galaxy
G1 (at redshift 0.867), and about 3.6 to 4.5 magnitudes fainter than
G1 depending on the wavelength. The photometric redshift of G2 is
consistent with that of G1 (although also consistent with a redshift
$\sim 4-5$). The inclusion of this faint satellite galaxy in the
model can explain the flux anomaly reasonably well, although the
satellite is required to be very flattened with an axis ratio of 8:1,
which may not be realistic.
This case highlights the potential roles that the luminous satellites
may play in the anomalous flux ratio problem. We note, however, that
numerical simulations by \citet{Dolag2008} showed that
star-dominated galaxies (not traced by dark matter only simulations)
appear to contribute only $\sim 10\%$ of the subhalo population in
clusters of galaxies. It is unclear however whether this
cluster-based result can be extrapolated to galaxy scales where
cooling is more efficient.  We plan to use semi-analytical galaxy
catalogues in the Aquarius simulations to address this issue more
quantitatively in subsequent work.

Substructures not only perturb the flux ratios, but also affect the image
positions. In Table \ref{tab:Aquarius_HalosCusp}, we show the maximum
perturbation of the deflection angle, $\alpha_{\rm sub, max}$, within the
central $2\arcsec \times 2\arcsec$ region, produced by all the subhaloes
within $r_{200}$. The maximum deviations range from a few milli-arcseconds to
$< 0.1$ arcseconds.  They may leave observable signatures on close pair images
such as that observed in MG2016+112 (\citealt{Koopmans2002, More2009}). We
find that most of these astrometric deviations are dominated by one large,
nearby subhalo. This clearly warrants further work in the near future.

\section*{Acknowledgements}
We thank Ian Browne, Neal Jackson, and Peter Schneider for useful
discussions. We also acknowledge an anonymous referee for 
constructive comments that improved the paper. DDX has been supported 
by a Dorothy Hodgkin fellowship for her postgraduate studies. 
LG acknowledges support from a STFC
advanced fellowship, one-hundred-talents program of the Chinese
Academy of Sciences (CAS) and the National basic research program of
China (973 program under grant No. 2009CB24901). SM acknowledges
travel support from the Humboldt Foundation and European Community's
Sixth Framework Marie Curie Research Training Network Programme,
contract number MRTN-CT-2004-505183 ``ANGLES''. GL is supported by
the Humboldt Foundation. The simulations for the Aquarius Project
were carried out at the Leibniz Computing Centre, Garching, Germany,
at the Computing Centre of the Max-Planck-Society in Garching, at
the Institute for Computational Cosmology in Durham, and on the
`STELLA' supercomputer of the LOFAR experiment at the University of
Groningen.

\clearpage

\clearpage

\clearpage

\clearpage

\clearpage

\clearpage

\clearpage

\clearpage

\clearpage
\begin{table*}
\centering
\caption{Substructure-lensing parameters of Aquarius haloes:}
\label{tab:Aquarius_HalosCusp}
\begin{minipage} {\textwidth}
\begin{tabular}[b]{c|c|c|c|c|c|c|c|c}\hline
~~~Halo Name ~~~ & ~~~$b_{\rm I}$~~~ &~~~ $q_3$~~~ &~~$S_0$~~~
&$f_{\rm sub, annu}$&$\alpha_{\rm sub, max}$
&$P(\Rcusp \ge 0.187)$ &$f_{\sigma}(\Delta\theta \le 90^{\circ})$& $M_{\rm sub, cr}$ \\
Projection & ($\arcsec$) & & ($\arcsec$) & (per cent) & ($\arcsec$) &
(per cent) & (per cent) & ($h^{-1}M_{\odot}$) \\\hline
Aq-A-2  \\
X-projection & 0.602 & 0.77 & 0.103 & 0.11 & 0.065 & 5.90 & 4.85 & $ 10^7 \Downarrow $ \\
Y-projection & 0.657 & 0.78 & 0.092 & 0.69 & 0.059 & 1.26 & 5.76 & $ 10^8 \Uparrow$  \\
Z-projection & 0.647 & 0.76 & 0.091 & 0.01 & 0.055 & 0.08 & 6.59 & $ 10^7 \Downarrow $  \\
\hline
Aq-B-2  \\
X-projection & 0.306 & 0.73 & 0.076 & 0.42 & 0.008 & 5.91 & 2.16 &  ---  \\
Y-projection & 0.408 & 0.91 & 0.071 & 0.48 & 0.007 & 64.13 & 0.41 & $ 10^8 \Downarrow $ \\
Z-projection & 0.286 & 0.64 & 0.080 & 0.09 & 0.012 & 0.09 & 0.77 & ---  \\
\hline
Aq-C-2 \\
X-projection & 0.846 & 0.91 & 0.085 & 0.13 & 0.015 & 19.09 & 1.07 & $ 10^8 \Downarrow $  \\
Y-projection & 0.571 & 0.60 & 0.107 & 0.06 & 0.002 & 13.98 & 21.41 & --- \\
Z-projection & 0.589 & 0.70 & 0.104 & 0.03 & 0.007 & 3.71 & 10.58 & $ 10^7 \Downarrow $ \\
\hline
Aq-D-2  \\
X-projection & 0.576 & 0.83 & 0.101 & 0.13 & 0.006 & 3.79 & 2.13 & $ 10^7 \Downarrow $ \\
Y-projection & 0.655 & 0.91 & 0.088 & 0.06 & 0.005 & 9.72 & 0.57 & $ 10^7 \Downarrow $ \\
Z-projection & 0.583 & 0.79 & 0.101 & 0.40 & 0.011 & 30.58 & 4.48 & $ 10^8 \Downarrow $ \\
\hline
Aq-E-2 \\
X-projection & 0.473 & 0.69 & 0.076 & 0.18 & 0.020 & 3.04 & 7.62 & $ 10^8 \Downarrow $  \\
Y-projection & 0.548 & 0.79 & 0.056 & 0.13 & 0.007 & 5.40 & 3.29 & $ 10^7 \Downarrow $ \\
Z-projection & 0.474 & 0.82 & 0.069 & 0.05 & 0.006 & 0.65 & 1.59 & $ 10^7 \Downarrow $ \\
\hline
Aq-F-2 \\
X-projection & 0.416 & 0.86 & 0.080 & 0.19 & 0.021 & 5.83 & 0.67 & $ 10^8 \Downarrow $  \\
Y-projection & 0.370 & 0.67 & 0.091 & 0.09 & 0.009 & 1.38 & 3.78 & --- \\
Z-projection & 0.435 & 0.85 & 0.088 & 0.22 & 0.012 & 6.74 & 0.78 & $ 10^8 \Downarrow $ \\
\hline
Aq-A-2 ($Z$ = 0.6) \\
X-projection & 0.568 & 0.71 & 0.079 & 0.11 & 0.008 & 0.60 & 7.75 & $ 10^8 \Downarrow $ \\
Y-projection & 0.731 & 0.89 & 0.054 & 0.70 & 0.022 & 55.81 & 1.74 & $ 10^8 \Downarrow $ \\
Z-projection & 0.592 & 0.69 & 0.082 & 0.33 & 0.083 & 7.56 & 12.00 & $ 10^8 \Downarrow $ \\
\hline
\end{tabular}
\\
Note: Cols (2-4): $b_{\rm I}$, $q_3$ and $S_0$, the Einstein radius,
axis ratio and core radius of the fitted isothermal ellipsoid 
(see eq. [\ref{eq:aboutrho}]); Col (5): $f_{\rm
  sub, annu}$ is the subhalo mass fraction within a $0.1\arcsec$-annulus
around the outer critical curve; Col (6): $\alpha_{\rm sub, max}$ is the
maximum magnitude in the projected central $2\arcsec\times 2\arcsec$ region of
the deflection angle due to all substructures within $r_{200}$,
usually found close to an individual subhalo;
Col (7): $P(\Rcusp \ge 0.187)$ is the probability (in per cent)
for sources with $\Delta\theta \le 90^{\circ}$ (defined as ``cusp sources'')
to have $\Rcusp \ge 0.187$, referred to as the
``cusp-caustic violation probability''; Col (8):
$f_{\sigma}(\Delta\theta \le 90^{\circ})$
is the cross-section fraction (as defined in eq. [\ref{eq:psigma}])
in the source plane for producing three
close images with opening angle $\Delta\theta \leq 90 ^{\circ}$; Col (9):
$M_{\rm sub, cr}$ is the critical subhalo mass that causes the strong violation
of the cusp-caustic relation. Arrows indicate ``above'' or ``below''.  Lenses
with naked cusps of the caustic always have low cusp-caustic
violation probability and are labelled as ``---''.\\
\end{minipage}
\end{table*}

\clearpage

\label{lastpage}

\end{document}